\documentclass[11pt,a4paper]{article}
\usepackage{amsfonts}       
\usepackage{amsmath}        
\usepackage{amssymb}
\usepackage{epsfig}
\usepackage{amsthm}
\usepackage{color}
\usepackage{amscd}
\usepackage{amstext}
\usepackage{slashed}
\usepackage{feynmp}
\newcommand{\Appendix}[1]{
\refstepcounter{section}
\begin{equation}gin{flushleft}
{\large\bf Appendix \thesection: #1}
\end{flushleft}}

\usepackage{fancybox}
\usepackage{color}

\usepackage{young}
\usepackage{color}
\usepackage{graphicx}

\usepackage[T1]{fontenc}
\usepackage[utf8]{inputenc}
\usepackage{authblk}

\title{ D3-D7 Holographic dual of a perturbed 3D CFT}
\author{ Hamid Omid, Gordon W. Semenoff}
\affil{\small  Department of Physics and Astronomy, University of British Columbia,\\
Vancouver, British Columbia, Canada V6T 1Z1}

\date{}

\begin{document}

  \maketitle

\abstract{An appropriately oriented D3-D7-brane system is the
  holographic dual of relativistic Fermions occupying a
  2+1-dimensional defect embedded in 3+1-dimensional spacetime. The
  Fermions interact via fields of ${\mathcal N}=4$ Yang-Mills theory
  in the 3+1-dimensional bulk.  Recently, using internal flux to
  stabilize the system in the probe $N_7<<N_3$ limit, a number of
  solutions which are dual to conformal field theories with Fermion
  content have been found.  We use holographic techniques to study
  perturbations of a particular one of the conformal field theories by
  relevant operators.  Generally, the response of a conformal field
  theory to such a perturbation grows and becomes nonperturbative at
  low energy scales.  We shall find that a perturbation which switches
  on a background magnetic field $B$ and Fermion mass $m$ induces a
  renormalization group flow that can be studied perturbatively in the
  limit of small $m^2/B$.  We solve the leading order explicitly.  We
  find that, for one particular value of internal flux, the system
  exhibits magnetic catalysis, the spontaneous breaking of chiral
  symmetry enhanced by the presence of the magnetic field.  In the
  process, we derive formulae predicting the Debye screening length of
  the Fermion-antiFermion plasma at finite density and the diamagnetic
  moment of the ground state of the Fermion system in the presence of
  a magnetic field.  }

\vskip 2cm


\section{Introduction}

 The AdS/CFT correspondence \cite{maldacena} offers the hope of
 direct, mathematically precise and systematically correctable study
 of the strong coupling limit of some quantum systems \cite{hartnoll}.
 Condensed matter physics in particular encounters a number of systems
 which exhibit quantum critical behavior and where the coupling can be
 argued to be strong.  In this Paper, we shall study the holographic
 dual of 2+1 dimensional quantum field theories with relativistic
 fermions.  Potential applications could be to condensed matter
 systems which have emergent relativistic 2+1-dimensional Fermions,
 examples of which are graphene \cite{Semenoff:1984dq}
 \cite{graphene}, topological insulators \cite{kane}, the D-wave state
 of high $T_c$ superconductors \cite{balents} and simulation of such
 systems on optical lattices \cite{dunne}.

The Coulomb force in graphene in particular is strong.  However, it
also violates the relativistic Lorentz symmetry of the free low energy
electrons. Our analysis in the following, being relativistic, only
applies if graphene finds a way to be relativistic even in the
presence of strong non-relativistic interactions. This could happen,
for example, if the relativistic theory is a conformal field theory
occurring at the infrared fixed point of a renormalization group flow.
There are some experimental indications that this could be the case.
However, there is little theoretical support for this idea, part of
the difficulty being the absence of reliable techniques for the strong
coupling regime.  We will not address this problem directly in this
paper. What has been done in previous work \cite{Davis:2011gi} is to
demonstrate that, at strong coupling, conformal field theories that
are viable candidates do indeed exist.  Here we shall examine some of
the properties of the strongly coupled conformal field theory.  In
particular, we shall be interested in the fate of the field theory
when it is perturbed by relevant operators.  Such perturbations, such
as turning on finite charge density, external magnetic fields or a
Fermion mass operator corresponding to sublattice asymmetric charge
density are very relevant to the physical properties of such systems.

In weakly coupled field theory, the propensity of gauge field mediated
interactions to form a chiral condensate and break chiral symmetry is
greatly enhanced by the presence of an external magnetic field.  This
phenomenon is called magnetic catalysis
\cite{klimenko}-\cite{Shovkovy:2012zn}.  The interesting question as
to whether it persists at strong coupling has been addressed in some
holographic models where it has indeed been found to occur,
particularly in holographic D3-D5 systems
\cite{Filev:2009xp}-\cite{Grignani:2012jh}.  However, it has
proven to be more elusive in the D3-D7 system \cite{Davis:2011gi} and
there are even cases of ``anti-catalysis'', suppression of a chiral
condensate by a magnetic field \cite{Preis:2010cq}\cite{Davis:2011am}.
Here, with our explicit perturbative solution of the holographic system,
we shall find that magnetic catalysis can indeed occur, but only for one special
value of a particular tuneable parameter.

An example of a top-down holographic construction of strongly
interacting 2+1-dimensional relativistic Fermions uses appropriately
oriented probe D7-branes in the $AdS_5\times S^5$ geometry that is
sourced by $N$ coincident D3-branes
\cite{rey}\cite{Myers:2008me}\cite{Bergman:2010gm}\cite{Davis:2011gi}.
The holographic construction begins with D7 and D3 branes oriented as
in Table 1 (\ref{dbranes}).
\begin{equation}\label{dbranes}
\boxed{\begin{array}{rcccccccccccl}
  & & x^0 & x^1 & x^2 & x^3 & x^4 & x^5 & x^6 & x^7 & x^8 & x^9 &\\
& D3 & \times & \times & \times & \times & & &  & & & & \\
& D7 & \times & \times & \times &  & \times  & \times & \times & \times & \times & &   \\
\end{array}}
\end{equation}
\begin{equation}
{\rm\bf Table~1:~D3-D7~orientation}
\nonumber
\end{equation}
The $N $ $D3$-branes and the $N_7$ $D7$ branes are extended in 2+1-spacetime
dimensions $(x^0, x^1, x^2)=(t,x,y)$ where they have $SO(2,1)$ Lorentz
symmetry. The lowest energy states of the 3-7 open strings are $N$
species of 2+1-dimensional 2-component Fermions.  The $x^9$ direction
is orthogonal to both the $D3$ and $D7$. The $D3$ and $D7$ can be
separated in that direction, introducing a bare mass for 3-7 strings.
For 2-component Fermions, a bare mass must violate parity. We will
discuss how parity is formulated in the D-brane construction shortly
and we will see that parity must be formulated to
change the sign of the separation of the D3 and D7 branes.

To apply holography, the limit where the number of $D3$-branes $N$ is
large is taken while holding the product of $N$ and the closed string
coupling constant $g_s$ fixed.  In the holographic duality, the closed
string coupling constant is related to the Yang-Mills coupling of the
bulk ${\mathcal N}=4$ supersymmetric Yang-Mills theory by $4\pi
g_s=g_{\rm YM}^2$. The quantity which is held fixed in the large $N$
limit is $4\pi g_sN=g_{YM}^2N\equiv \lambda$, the 't Hooft coupling of
the gauge theory.  Then, the D3-branes are replaced by the
$AdS_5\times S^5$ geometry. The radii of curvature of the $AdS_5$ and
$S^5$ are $L=\lambda^{\tfrac{1}{4}}\sqrt{\alpha'}$. The $D7$-branes
are treated as probes and the dynamical problem is to find their
embedding in $AdS_5\times S^5$.

This D3-D7 configuration has a unique feature that it is
non-supersymmetric, but is free of tachyons and the only low energy
modes of the D3-D7 open strings are Fermions. As a consequence, the
decoupling limit produces a system which at weak coupling contains
only chiral Fermions.  Being a non-supersymmetric configuration, the
D3 and D7-branes repel each other.  This shows up as an instability
that appears when one attempts to embed the D7-brane in $AdS_5\times
S^5$.  Fluctuations of the geometry violate the Breitenholder-Freedman
bound in the large AdS radius regime.  This instability can be fixed
by introducing flux of the world-volume gauge fields of the D7-brane,
either an instanton bundle \cite{Myers:2008me} or U(1) magnetic
magnetic monopole fluxes \cite{Bergman:2010gm}.  In the latter case,
which is the one we will focus on in this paper, the four dimensions
of the D7 world-volume which are embedded in $S^5$ are taken as two
2-spheres, $S^2$ and $\tilde S^2$, and each 2-sphere has a number
$n_D$ and $\tilde n_D$ units of Dirac magnetic monopole flux.  It was
shown in reference \cite{Bergman:2010gm} that the latter configuration
is stable, at least to small fluctuations  if either $n_D$ or $\tilde
n_D$ is large enough. The decoupling limit of the D3-D7 brane intersection which
produces a D7-brane with geometry $AdS_4\times S^2\times S^2$ is discussed
in section 2 of reference \cite{Bergman:2010gm} and we refer the reader to their exposition for
the details.

\begin{figure}
 ~\includegraphics[scale=.5]{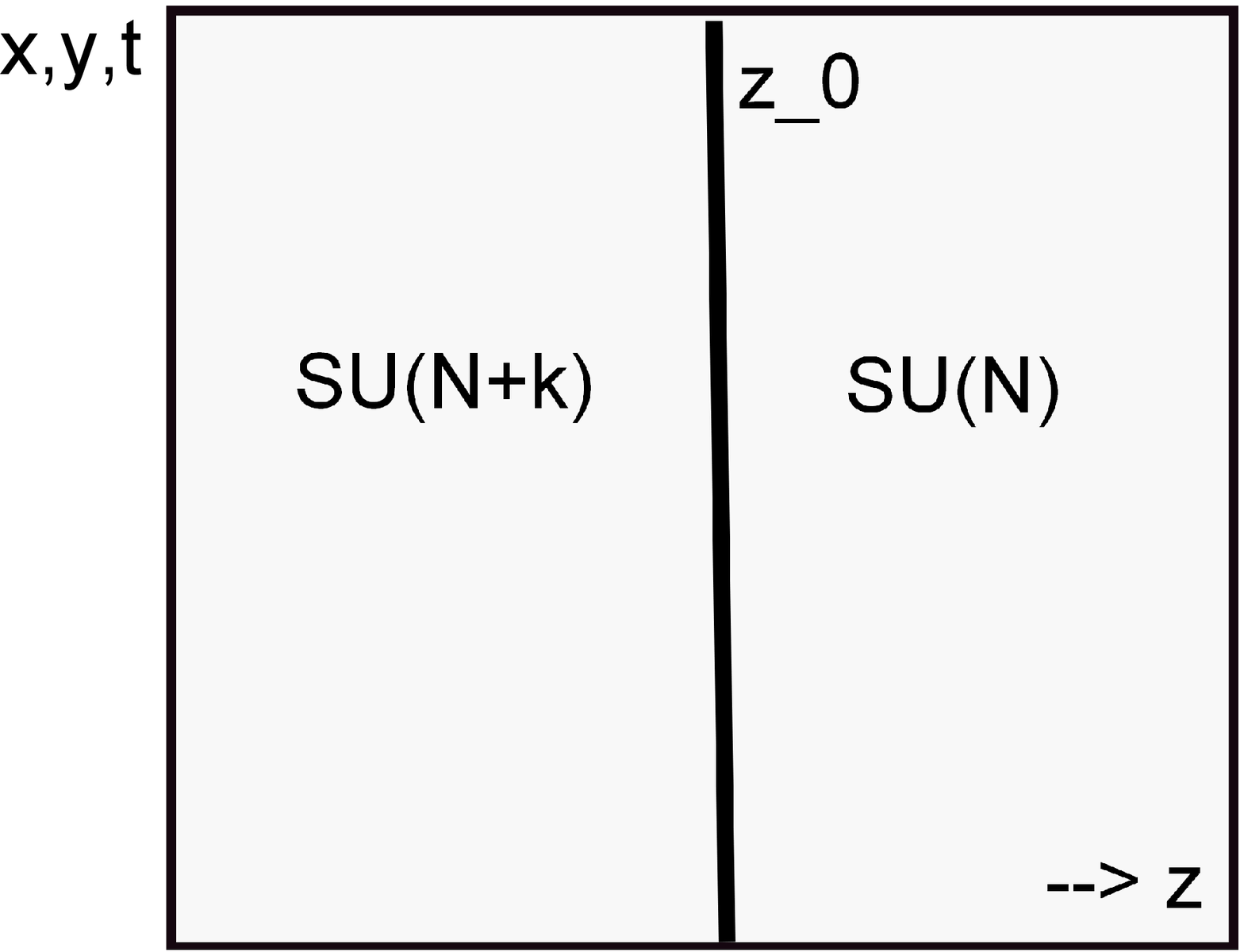}\\
\begin{caption} {{\footnotesize Defect conformal
field theory: Fermions are constrained to occupy a plane, denoted
by the vertical line through the center of the diagram.  This plane
divides the three dimensional space into two  regions which
are occupied by four dimensional conformal ${\cal N}=4$ supersymmetric
Yang-Mills theories with different gauge groups. The conformal field theory
has three tunable parameters, $N$, $N + k$ and the Yang-Mills coupling constants. The holographic
description describes the planar limit of this theory where the 't Hooft coupling is tuned to be
large. The remaining parameter is $k = n_{D}\tilde n_D$.
\label{cft}}}\end{caption}
\end{figure}

The quantum field theory which is dual to the D3-D7 system is a defect
field theory consisting of Fermions confined to a 2-dimensional plane
which separates 3-dimensional space into two regions as depicted in
figure \ref{cft}. The 3+1-dimensional bulk is occupied by ${\cal N}=4$ supersymmetric Yang-Mills theory
with gauge group SU(N) on one side of the defect and gauge group SU(N+k) on the
other side of the defect. There are $N_7$ species of 2-component spinors of
the 2+1-dimensional Lorentz group $SO(2,1)$, living on the defect.  In the conformal
invariant solution this symmetry is extended to the 2+1-dimensional
conformal group $SO(3,2)$.  The Fermions transform in the fundamental
representation of a global $U(N_7)$ symmetry.  Since we consider no
processes which use the non-abelian nature of $U(N_7)$, for simplicity
we will take $N_7=1$ (and remember that to apply to graphene, we need
$N_7=4$ to produce the correct flavor symmetry).  The Fermions also
transform in the fundamental representation of the gauge group of
${\cal N}=4$ supersymmetric Yang-Mills theory which inhabits the
3+1-dimensional bulk of the space-time.  The bulk Yang-Mills theory
has different gauge groups on each side of the defect, as shown in
figure \ref{cft}.  This is a result of the fact that, the D7-branes
with internal fluxes which we shall use can be described as D7-branes
with $n_D\tilde n_D$ D3-branes dissolved into their worldvolumes.  The
D7-brane then forms a boundary between regions with different numbers
of D3-branes and therefore different amounts of Ramond-Ramond 4-form
flux, thus different ranks of the gauge group in the field theory
dual.

 We shall be interested in field theories which become parity and
 charge conjugation invariant in their high energy limit.  This is
 what is expected in a class of condensed matter theories where the
 underlying dynamics is parity and particle-hole symmetric.  These
 symmetries can be broken by deformations like external magnetic
 field, chemical potential or parity violating mass terms which are
 irrelevant in the ultraviolet limit but can be important to the
 infrared properties of the theory.  On the D-brane side, imposing
 parity and charge conjugation symmetry will involve taking the
 appropriate boundary condition for the embedding of the D7-brane in
 $AdS_5\times S^5$ as well as setting the fluxes equal, $n_D=\tilde
 n_D$.

The paper is organized as follows: In Section 2 we review the
embedding of the probe D7-brane in $AdS_5\times S^5$. In Section 3 we
discuss the properties of the conformally invariant solution of the
embedding problem.  In Section 4 we discuss the solution with a
chemical potential and charge density.  In particular, we derive
expressions for the Debye screening length at strong coupling, as
functions of chemical potential and of density.  In Section 5 we
examine the same system with an external magnetic field in the special
case that the charge density is tuned to zero.  We find a simple
expression for the diamagnetic moment of the system.  We solve the
embedding equation perturbatively in the ratio of condensate to
magnetic field. We find a relationship between the mass $m$ and the
chiral condensate $c$ to linear order in $m^2/B$ in equation
(\ref{result}). In Section 7 we show that turning on an infinitesimal
charge density can also be taken into account perturbatively and we
write the embedding equation to leading order in the filling fraction
$\rho/B$.

\section{D7-brane}

In the limit where the string theory is classical, the problem of
embedding a D7-brane in the $AdS_5\times S^5$ geometry reduces to that
of finding an extremum of the Dirac-Born-Infeld and Wess-Zumino
actions,
\begin{align}\label{dbi}
S=\frac{ T_7}{g_s} \int d^8\sigma\left[- \sqrt{-\det( g+2\pi\alpha' F)} +\frac{(2\pi\alpha')^2}{2}
C^{(4)}\wedge F\wedge F\right] 
\end{align}
where $g_s$ is the closed string coupling constant, which is related
to the ${\mathcal N}=4$ Yang-Mills coupling by $4\pi g_s =g_{YM}^2$,
$\{\sigma_0,\sigma_1,\ldots,\sigma_7\}$ are the coordinates of the
D7-brane world-volume, $g_{ab}(\sigma)$ is the induced metric,
$C^{(4)}$ is the 4-form of the $AdS_5\times S^5$ background, $F$ is
the world-volume gauge field and the D7 brane tension is
\begin{align}\label{t7}
T_7=\frac{1}{(2\pi)^7{\alpha'}^4}
\end{align}
We shall work with coordinates where the metric of the background
space $AdS_5\times S^5$ is
\begin{align}\label{ads5metric}
ds^2= L^2&\left[ r^2 (-dt^2+dx^2+dy^2+dx^2) +
  \frac{dr^2}{r^2}+\right. \nonumber
  \\ &\left. +d\psi^2+\cos^2\psi(d\theta^2+\sin^2\theta d\phi^2)+
  \sin^2\psi (d\tilde\theta^2+\sin^2\tilde\theta d\tilde\phi^2)\right]
\end{align}
Here, $(t,x,y,z,r)$ are coordinates of the Poincare patch of
$AdS_5$. In our notation, and in natural units $\hbar=1$ and $c=1$,
$r$ has the dimension of inverse length and $(x,y,z,t)$ have
dimensions of length. $L$ is the radius of curvature. The boundary of
$AdS_5$ is located at $r\to\infty$ and the Poincare horizon at $r\to
0$.  The 5-sphere is represented by two unit 2-spheres, $S^2$ with
polar coordinates $(\theta,\phi)$ and $\tilde S^2$ with coordinates
$(\tilde\theta,\tilde\phi)$.  The 2-spheres are fibered over the
interval $\psi\in[0,\tfrac{\pi}{2}]$.  The Ramond-Ramond 4-form is
\begin{align}\label{4form}
C^{(4)}=  L^4r^4dt\wedge dx\wedge dy\wedge dz+L^4\frac{c(\psi)}{2}d\cos\theta\wedge d\phi
\wedge d\cos\tilde\theta\wedge d\tilde\phi
\end{align}
with $c(\psi)$ obeying the equation
\begin{align}\label{c}
\partial_\psi c(\psi)=8\sin^2\psi\cos^2\psi
\end{align}
The dynamical variables are the ten functions of eight world-volume
coordinates which embed the D7-brane in $AdS_5\times S^5$,
$$
\{ x(\sigma),y(\sigma),z(\sigma),t(\sigma),r(\sigma),\psi(\sigma),\theta(\sigma),\phi(\sigma),
\tilde\theta(\sigma),\tilde\phi(\sigma)\}
$$
as well as the eight worldvolume gauge fields
$$
\{A_0(\sigma),A_1(\sigma),\ldots,A_7(\sigma)\}
$$

Parity and charge conjugation are important symmetries for the class
of quantum field theory systems that we are interested in. For
example, graphene is certainly parity invariant and, to a good
approximation, it has particle-hole symmetry.  We would expect to
model it using field theories that have parity and charge conjugation
symmetry. The string theory dual of such a field theory should also
have these symmetries.  We should therefore make sure that the problem
of finding a minimum of the action (\ref{dbi}) itself is symmetric.
Parity in two dimensions is a reflection of one of the spatial
coordinates, $x\to -x$.  This symmetry is usually broken by
Wess-Zumino terms, which are\footnote{Note that we have presented the
  second Wess-Zumino term (\ref{cs2}) in a form that is integrated by
  parts. This was to avoid specifying the integration constant in
  $c(\psi)$ which would be obtained by integrating the expression in
  (\ref{c}). Specifying that integration constant (which we shall do
  in the following) is regarded as string theory gauge fixing and
  physical quantities should not depend on it. We can think of our
  integration by parts in (\ref{cs2}) as equivalent to adding a
  surface term to the Wess-Zumino term in order to restore this gauge
  invariance.}
\begin{align}
\label{cs1}
& \int d^6\sigma \epsilon^{\mu_0\mu_1\ldots\mu_7} \partial_{\mu_0}t(\sigma) \partial_{\mu_1}x(\sigma) \partial_{\mu_2} y(\sigma)\partial_{\mu_3}z(\sigma)  r^4(\sigma) ~\partial_{\mu_4}A_{\mu_5}(\sigma) \partial_{\mu_6} A_{\mu_7}(\sigma)
\\ \label{cs2}
& \int d^6\sigma \epsilon^{\mu_0\mu_1\ldots\mu_7} \partial_{\mu_0}\cos\theta(\sigma)\partial_{\mu_1}\phi(\sigma)
 ~\partial_{\mu_2} \cos\tilde\theta(\sigma)\partial_{\mu_3}\tilde\phi(\sigma)
 c'(\psi)\partial_{\mu_4} \psi(\sigma)  A_{\mu_5}(\sigma)  \partial_{\mu_6} A_{\mu_7}(\sigma)
\end{align}
Indeed, (\ref{cs1}) changes sign when $x\to -x$.  We must therefore
compensate the sign change by another change of the world-sheet
variables, say $\sigma_1\to -\sigma_1$.  However, now (\ref{cs1}) is
invariant but (\ref{cs2}) changes sign and is not invariant.  We can
also make it invariant by another change of variables,
$(\psi,\theta,\phi,\tilde\theta,\tilde\phi)\to (\tfrac{\pi}{2}-\psi,
\tilde\theta,\tilde\phi,\theta,\phi)$. (Note that $c'(\psi)$ is
invariant and $\partial\psi$ changes sign under this transformation.)
Then, both of the Wess-Zumino terms are invariant, as is the
Dirac-Born-Infeld action. In summary, our parity transformation is the
replacement
 \begin{align}
{\rm P}:~&\{ x'(\sigma'),y'(\sigma'),z'(\sigma'),t'(\sigma'),r'(\sigma'),\psi'(\sigma'),\theta'(\sigma'),\phi'(\sigma'),
\tilde\theta'(\sigma'),\tilde\phi'(\sigma')\}= \nonumber \\
&=\{ -x(\sigma),y(\sigma),z(\sigma),t(\sigma),r(\sigma),\tfrac{\pi}{2}-\psi(\sigma),\tilde\theta(\sigma),\tilde\phi(\sigma),
\theta(\sigma),\phi(\sigma)\}   \label{parity1} \\
A_\mu'(\sigma')&=\{A_0(\sigma),-A_1(\sigma),A_2(\sigma),\ldots,A_7(\sigma)\} \label{parity2}\\
\sigma'&=\{\sigma_0,-\sigma_1,\sigma_2,\ldots,\sigma_7\}  \label{parity3}
\end{align}
Charge conjugation symmetry (C) is the replacement $A_\mu\to -A_\mu$
and the Wess-Zumino terms are invariant.  However, we shall introduce
a background field
\begin{align}\label{backgroundmagneticfield}
F_0= \frac{L^2}{2\pi\alpha'}\frac{f}{2}\left(d\cos\theta\wedge d\phi+d\cos\tilde\theta\wedge d\tilde\phi\right)
\end{align}
Here, $f$ is the strength of the monopole bundle\footnote{A monopole
  bundle has quantized flux.  Here the number of quanta is very large
  in the strong coupling limit $2\pi n_D=\sqrt{\lambda}f$, so that it
  is to a good approximation a continuously variable parameter.  }
which is needed to stabilize the system. Note that, to be invariant
under parity, (\ref{parity1})-(\ref{parity3}), the fluxes on the
2-spheres have to be equal, $n_D=\tilde n_D$. This is seen most
clearly by noting that, in (\ref{parity1})-(\ref{parity3}), the
transformation interchanges the 2-spheres.  The background field
breaks a symmetry if the two components in the total field
$F=F_0+\delta F$ transform differently. Under C, $\delta F\to -\delta
F$.  We shall need a definition of C so that $F_0\to -F_0$.
Such a transformation is
  \begin{align}
{\rm C}:~&\{ x'(\sigma),y'(\sigma),z'(\sigma),t'(\sigma),r'(\sigma),\psi'(\sigma),\theta'(\sigma),\phi'(\sigma),
\tilde\theta'(\sigma),\tilde\phi'(\sigma)\}= \nonumber\\
&=\{ x(\sigma),y(\sigma),z(\sigma),t(\sigma),r(\sigma), \psi(\sigma), \theta(\sigma),2\pi-\phi(\sigma),
\tilde\theta(\sigma),2\pi-\tilde\phi(\sigma)\}   \label{c1} \\
A_\mu'(\sigma)&=\{-A_0(\sigma),-A_1(\sigma),-A_2(\sigma),\ldots,-A_7(\sigma)\} \label{c2}
\end{align}
Then (\ref{cs1}) and (\ref{cs2}) are invariant and
(\ref{backgroundmagneticfield}) transforms covariantly under our
definition of C. We have established that the mathematical problem of
finding the D7 embedding has the discrete symmetries which will appear
as C and P for the 2+1-dimensional Fermions. In the case of $C$, this
is clear, the worldsheet gauge field $A_\mu$ is dual to a conserved
U(1) current $j^a=\bar\psi\gamma^a\psi$ for the Fermions and $A\to -A$
corresponds to $j^a\to -j^a$, which is implemented by the usual C
transformation of Fermi fields.  For parity, this is also the case,
the Fermion kinetic term is covariant under $x\to -x$.  However a
2-component Fermion mass operator $\bar\psi\psi$ changes sign under
this parity transformation, so parity takes the place of chiral
symmetry in 2+1-dimensions in that it protects the masslessness of
Fermions. Indeed, the geometric argument of reference
\cite{Bergman:2010gm} explains that the deviation of the angle $\psi$
from its symmetric value $\tfrac{\pi}{4}$ corresponds to a separation
of the D3 and D7 branes and a mass for the D3-D7 strings.  At the same
time, this deviation would violate parity.

Our Ans\"atz for a solution will describe a D7-brane covering the
whole range of $(t,x,y,r,\theta,\phi,\tilde\theta,\tilde\phi)$ and it
is for the most part determined by symmetry,
\begin{align}\label{ansatz}
t=\sigma_0,x=\sigma_1,y=\sigma_2,z=z(\sigma_3),r=\sigma_3,
\theta=\sigma_4,\phi=\sigma_5,\tilde\theta=\sigma_6,\tilde\phi=\sigma_7,\psi=\psi(\sigma_3)
\end{align}
we will denote the coordinate $\sigma_3$ by $r$. The two unknown functions in this embedding are
then $z(r)$ and $\psi(r)$.
Our Ans\"atz for the world-volume gauge fields is
\begin{align}\label{magneticfields}
F= \frac{L^2}{2\pi\alpha'} a'(r)dr\wedge dt+
\frac{L^2}{2\pi\alpha'}bdx\wedge dy+
\frac{L^2}{2\pi\alpha'}\frac{f}{2}\left(d\cos\theta\wedge
d\phi+d\cos\tilde\theta\wedge d\tilde\phi\right)
\end{align}
where $b$ is a constant magnetic field which is proportional to a
constant magnetic field in the dual field theory and $a(r)$ is the
temporal world-volume gauge field which must be non-zero in order to
have a uniform charge density in the field theory dual.\footnote{$b$
  and $q$ are related to the physical magnetic field and charge
  density as $ b=\frac{2\pi}{\sqrt{\lambda}}B$, $q=
  \frac{8\pi^4}{\lambda N }\frac{\sqrt{\lambda}}{2\pi}\rho $ so that
$$
\frac{q}{b}= \pi\nu ~~,~~\nu = \frac{1}{N} \frac{2\pi \rho}{B}
$$
where the dimensionless parameter $\nu$ is the filling fraction of $N$ degenerate Landau levels.}

With the Ans\"atz (\ref{ansatz}) and (\ref{magneticfields}), the
D7-brane world-volume metric is
\begin{align}\label{D7metric}
ds^2= L^2\left[ r^2
  (-dt^2+dx^2+dy^2)+\frac{dr^2}{r^2}(1+r^2{\psi'}^2+r^4{z'}^2)+\right. \nonumber
  \\ \left. +\cos^2\psi(d\theta^2+\sin^2\theta d\phi^2)
  +\sin^2\psi(d\tilde\theta^2+\sin^2\tilde\theta d\tilde\phi^2)\right]
\end{align}
where prime denotes derivative by $r$.
The Lagrangian is
\begin{align}\label{ansatzaction}
\frac{\mathcal L}{\mathcal N}= -
\sqrt{(f^2+4\cos^4\psi)(f^2+4\sin^4\psi)(b^2+r^4)(1+r^2{\psi'}^2+r^4{z'}^2)-{a'}^2}
\nonumber \\ +f^2 r^4z'+2abc'(\psi)\psi'
\end{align}
where, using (\ref{t7}),
\begin{align}\label{N}
{\mathcal N}=\frac{(2\pi)^2 T_7 L^8 }{g_s}V_{2+1}=\frac{\lambda  N}{8\pi^4}V_{2+1}
\end{align}
The factor of $(2\pi)^2$ in the numerator comes from half of the
volume of the unit 2-spheres (the other factors of 2 are still in the
action).  Since nothing depends on $(x,y,t)$, the integral over these
coordinates produces the volume factor $V_{2+1}$ which appears in
${\cal N}$.  The Wess-Zumino term gives a source for $z(r)$, so that,
as long as the flux $f$ is nonzero, $z(r)$ will be non-zero and
$r$-dependent.  Note that, since $V_{2+1}$ has dimension of length
cubed and $r$ has the dimension of inverse length, the integral of
(\ref{ansatzaction}) over $r$ will be dimensionless, as it should be.

Now, we must solve the equations of motion for the functions
$\psi(r)$, $a(r)$ and $z(r)$ which result from the Lagrangian
(\ref{ansatzaction}) and the variational principle. Since the
Lagrangian depends only on their derivatives and not on the variables
$a(r)$ and $z(r)$ themselves, $a(r)$ and $z(r)$ are cyclic variables
and they can be eliminated using their equations of motion,
\begin{align}\label{const1}
&\frac{d}{dr}\frac{\delta S}{\delta z'(r)}=0~~\to~~
  \frac{\sqrt{(f^2+4\cos^4\psi)(f^2+4\sin^4\psi)(b^2+r^4)}r^4z'}{\sqrt{1+r^2{\psi'}^2+r^4{z'}^2-{a'}^2}}-
  f^2 r^4 =p_z \\ \label{const2} &\frac{d}{dr}\frac{\delta S}{\delta
    a'(r)}=0~~\to~~
  \frac{\sqrt{(f^2+4\cos^4\psi)(f^2+4\sin^4\psi)(b^2+r^4)} {a'}
  }{\sqrt{1+r^2{\psi'}^2+r^4{z'}^2-{a'}^2}}-2bc=-q
\end{align}
where $ p_z$ and $q$ are constants of integration. $p_z$ can be
interpreted as being proportional to the pressure in the $z$-direction
and, by translation invariance, for a single brane, we would expect
that it would be zero if the brane is not accelerating\footnote{The
  parameter $p_z$ becomes important when there is more than one brane,
  as the branes interact with each other and a pressure is required to
  hold them together, or apart, depending on their relative
  orientations \cite{Davis:2011am}.  With a single brane, because of
  translation invariance, this pressure must vanish.}.  It is also
clear that, if the brane is to reach the Poincare horizon at $r\to 0$,
equation (\ref{const1}) will make sense there only if we set $p_z=0$.
Also, to get (\ref{const2}) we have integrated $2bc'$ to get the term
$2bc(\psi)$ in the equation.  We have taken the integration constant
so that $c(\tfrac{\pi}{2}-\psi)=-c(\psi)$, so that it has the correct
transformation property under P,
\begin{align}\label{c1}
c(\psi)=\psi-\tfrac{\pi}{4}-\tfrac{1}{4}\sin4\psi
\end{align}
The overall constant of integration in (\ref{c1}) is, of course, a
superstring gauge choice.  Other choices would give equivalent
results, but would alter what we mean by total charge density and
would complicate the parity transformation law.  Here, for simplicity,
we make the choice given in (\ref{c1}).  The integration constant $q$
is then proportional to the total charge density in the field theory
dual. Moreover, we can solve for $z'$ and $a'$,
\begin{align}
\label{equationforzp}
&z'=\frac{
  f^2\sqrt{1+r^2{\psi'}^2}}{\sqrt{(f^2+4\cos^4\psi)(f^2+4\sin^4\psi)
    (b^2+r^4)-f^4r^4+(q-2bc)^2}}
\\ &a'=\frac{(2bc-q)\sqrt{1+r^2{\psi'}^2}}{\sqrt{(f^2+4\cos^4\psi)(f^2+4\sin^4\psi)
    (b^2+r^4)-f^4r^4+(q-2bc)^2}} \label{aprime}
\end{align}
We must then use the Legendre transformation
$$
{\mathcal R}= S-\int a'(r)\frac{\partial L}{\partial a'(r)}-\int z'(r)\frac{\partial L}{\partial z'(r)}
$$ to eliminate $z'$ and $a'$.  We obtain the Routhian
\begin{align}
{\mathcal R}={\mathcal N} \int d r
\sqrt{(f^2+4\cos^4\psi)(f^2+4\sin^4\psi)
  (b^2+r^4)-f^4r^4+(q-2bc)^2}\sqrt{1+r^2{\psi'}^2}
\label{routhian1}
\end{align}
which must now be used to find an equation of motion for $\psi(r)$.
Applying the Euler-Lagrange equation to the Routhian (\ref{routhian1})
yields the equation of motion
\begin{align}
\frac{\ddot\psi}{1+\dot\psi^2}+\dot\psi\left[
  1+\frac{2r^4[(f^2+4\cos^4\psi)(f^2+4\sin^4\psi) -f^4]
  }{(f^2+4\cos^4\psi)(f^2+4\sin^4\psi)
    (b^2+r^4)-f^4r^4+(q-2bc)^2}\right] + \nonumber
\\ +\frac{4\sin2\psi\left[[f^2 ( b^2+r^4)-r^4\sin^22\psi] \cos2\psi +
    bq\sin2\psi -2b^2(\psi-\pi/4)\sin2\psi \right]
}{(f^2+4\cos^4\psi)(f^2+4\sin^4\psi) (b^2+r^4)-f^4r^4+(q-2bc)^2}=0
\label{equationforpsi}\end{align}
where the overdot is the logarithmic derivative
$\dot\psi=r\frac{d}{dr}\psi$.  In the next few Sections, we will
discuss some of the solutions of this equation.

\section{Conformal field theory}
\label{sec:cft}

Let us first review the relevant solutions for the D7-brane geometry
when the external magnetic field and the charge density vanish:
$b=0,q=0$.  These are described in references \cite{Davis:2011gi} and
\cite{Bergman:2010gm}-\cite{Jokela:2012vn}. The equation for $\psi(r)$
is
\begin{align}
\frac{\ddot\psi}{1+\dot\psi^2}+3\dot\psi +\frac{2
  \sin4\psi(f^2-\sin^22\psi)}{ (f^2+4\cos^4\psi)(f^2+4\sin^4\psi) -f^4
}=0
\label{equationforpsiconformal}
\end{align}
In the large $r$ regime, we require that the angle approaches the
parity symmetric solution, $\psi\to \tfrac{\pi}{4}$.  We also note
that the constant angle $\psi=\tfrac{\pi}{4}$ is an exact solution of
equation (\ref{equationforpsiconformal}).  Linearizing about that
solution yields the differential equation
\begin{equation}
\left(r\frac{d}{dr}\right)^2(\psi-\tfrac{\pi}{4})+3r\frac{d}{dr}(\psi-\tfrac{\pi}{4})+
\frac{8(1-f^2)}{(f^2+1)^2-f^4}(\psi-\tfrac{\pi}{4})=0
\end{equation}
which is solved by
\begin{equation}\label{fluctuations}
\psi(r)=\tfrac{\pi}{4}+\frac{m}{r^{\Delta_-}}+\frac{c}{r^{\Delta_+}}+\ldots
\end{equation}
where
\begin{equation}\label{power1}
\Delta_\pm = \frac{3}{2}\pm \frac{1}{2}\sqrt{9-\frac{32(1-f^2)}{(f^2+1)^2-f^4}}
\end{equation}
The argument of the square root is non-negative and the exponents are
real numbers if $f^2$ is large enough,
\begin{equation}
f^2\geq\frac{23}{50}
\end{equation}
This is the parameter regime where mass term in the linearized
equation (\ref{fluctuations}) does not violate the
Breitenholder-Freedman bound. In the regime $\frac{23}{50}\leq f^2\leq
1$, both $\Delta_+$ and $\Delta_-$ are positive, so that both solutions in (\ref{fluctuations})
go to zero at $r\to\infty$. Moreover, when they are both positive, both solutions diverge at $r\to 0$.
Thus we see that there is no small deviation from the solution $\psi=\frac{\pi}{4}$ which is itself
a solution.  For this reason we call $\psi=\frac{\pi}{4}$ an isolated solution.

Let us discuss this in a little more detail.  Regardless of its
behavior at $r>0$, if $\psi(r)$ is to have a ``normalizable mode''
which remains finite at $r\to 0$, for small values of $r$, the
solution must also go to a zero of the last term in
(\ref{equationforpsiconformal}).  When $\frac{23}{50}< f^2<1$, there
are three zeros of the last term in (\ref{equationforpsiconformal}),
$\psi=0,\tfrac{\pi}{4},\tfrac{1}{2}\arcsin f$.  The exponents of the
linearized equation in the region $r\to 0$, in each of the three cases
are
\begin{align}
\psi
&=0+\frac{\kappa_1}{r^{\gamma_-}}+\frac{\kappa_2}{r^{\gamma_+}}+\ldots~~,~&\gamma_-=1~,~\gamma_-=2 \label{asymptotic1}
\\ \psi&=\frac{\pi}{4}+\frac{\kappa_1}{r^{\gamma_-}}+\frac{\kappa_2}{r^{\gamma_+}}+\ldots~~,~&\gamma_{\pm}=\frac{3}{2}\pm
\frac{1}{2}\sqrt{9-
  \frac{32(1-f^2)}{(f^2+1)^2-f^2}} \label{asymptotic2}\\ \psi&=\tfrac{1}{2}\arcsin
f+\frac{\kappa_1}{r^{\gamma_-}}+\frac{\kappa_2}{r^{\gamma_+}}+\ldots~~,~&\gamma_{\pm}=\frac{3}{2}\pm
\frac{3}{2}\sqrt{1+\frac{64}{9}\frac{1-f^2}{4-f^2} }
  \label{asymptotic3}
\end{align}
Note that the exponent  for fluctuations about the  $\psi\to \frac{pi}{4}$ asymptotic which is
given in (\ref{asymptotic2}) for the small $r$ regime is identical to the one in the large $r$ regime given in (\ref{power1}).
The first solution $\psi=0$ would be acceptable only if both
$\kappa_1$ and $\kappa_2$ are zero.  The solution would then
necessarily be a constant and violates the boundary condition at large
$r$, where $\psi$ should go to $\frac{\pi}{4}$.  If the boundary
condition at infinity were different, so that the constant $\psi=0$
everywhere were a solution, this solution would describe a D5-brane
with flux in this geometry \cite{Grignani:2012jh}.  This is an
interesting possibility, but not the one that we need here.

The second solution (\ref{asymptotic2}) is the isolated
$\psi=\tfrac{\pi}{4}$ solution. It is constant and isolated for the
reasons that we have discussed above and it is the only solution that
does not violate parity.

The third small $r$ solution is $\psi(r\to 0)\to\tfrac{1}{2}\arcsin f$
which is not isolated -- since $\gamma_-$ is negative and $\gamma_+$ is positive, $\kappa_1$ is
allowed to be nonzero and we need $\kappa_2=0$ in order to have good
behavior at small $r$.  At large $r$, both constants $c$ and $m$ in
equation (\ref{fluctuations}) are allowed to be nonzero, so the
asymptotic behavior of the solutions is characterized by three nonzero
constants, $\kappa_1$, $m$ and $c$.  Generically, for a fixed value of
one of the constants, for example, $m$, both $c$ and $\kappa_1$ must be tuned to
obtain $\kappa_2=0$.  Then, $c$ and $\kappa_1$ are functions of $m$.
This means that we have a 1-parameter family of solutions
parameterized by $m$. This parameter is the holographic dual of a
parity violating Fermion mass term in the quantum field theory.
The $\psi=\tfrac{\pi}{4}$ solution has a relevant parity violating
operator with conformal dimension
$$
\Delta_+ = \frac{3}{2}+ \frac{1}{2}\sqrt{9-\frac{32(1-f^2)}{(f^2+1)^2-f^4}}
$$ which can be introduced and which has coupling constant dual to the
parameter $m$ and expectation value dual to the parameter
$c$.\footnote{In an alternative quantization, if $\Delta_+$ and
  $\Delta_-$ are both greater than $\tfrac{1}{2}$, the unitarity bound
  for conformal dimensions of operators in d=2+1, the roles $m$ and
  $c$ can be interchanged \cite{Klebanov:1999tb}.}  Turning on this
operator preserves charge conjugation symmetry but breaks parity
symmetry and it corresponds to making $m$ nonzero.  Once $m$ is
nonzero, no matter how small, the solution will evolve to the same
value, $\psi=\tfrac{1}{2}\arcsin f$ at $r=0$.  This can be
significantly different from the $\psi=\tfrac{\pi}{4}$ of the parity
invariant solution.  This has the interpretation of a renormalization
group flow driven by a relevant operator.  Note that, when $f^2=1$,
the exponents are $(\Delta_-,\Delta_+)=(0,3)$.  In the dual field
theory this is interpreted as having the Fermion mass operator with
dimension 3, that is, exactly marginal. In this case the angle goes to
$\tfrac{1}{2}\arcsin (1)\to\tfrac{\pi}{4}$ at small $r$. A numerical
solution of (\ref{equationforpsiconformal}) is easy to find.  Plots of
numerical solutions exhibiting this behavior were given in
reference~\cite{Davis:2011gi}.

\section{Conformal field theory with charge density}

We note from the Lagrangian (\ref{ansatzaction}) that it contains a
cubic term $abc'(\psi)\psi'$.  Remember that $a$ is $C$ and $CP$
violating, $b$ is P and C violating and $\psi'$ is P and CP violating.
This implies that turning on any two of these fields will violate all
three of the symmetries $P$, $C$ and $CP$ and, in the equation of
motion, turning on two  will produce a source for the third. In the following
sections we will discuss the situation when all three are turned on.
Before that, let us examine what happens when only one of the three is
turned on. In the above, we have already discussed one of the cases,
the situation when $a$ and $b$ are turned off but $\psi$ could be a
function of $r$.  That solution violated parity but preserved charge
conjugation invariance.  In this section, we will keep $\psi$ a
constant, set $b$ to zero and turn on $q$.  This solution violates
charge conjugation invariance, in that there is a fixed non-zero
charge density, but it is invariant under parity.  What we obtain is a
sector of the conformal field theory where the charge density is fixed
to a particular value proportional to $q$.  The equation for $\psi(r)$
is gotten from (\ref{equationforpsi}) by setting $b=0$,
\begin{align}
\frac{\ddot\psi}{1+\dot\psi^2}+\dot\psi\left[ 1+\frac{2[r^4(f^2+4\cos^4\psi)(f^2+4\sin^4\psi)
-f^4r^4]}{r^4(f^2+4\cos^4\psi)(f^2+4\sin^4\psi)
-f^4r^4+q^2}\right] + \nonumber \\
+\frac{2r^4\sin4\psi(f^2-\sin^22\psi)}{r^4(f^2+4\cos^4\psi)(f^2+4\sin^4\psi)  -f^4r^4+ q^2}=0
\label{equationforpsinob1}\end{align}
We see that $\psi=\frac{\pi}{4}$ is a solution of this equation.  The
remaining embedding functions can be found from by integrating the
expressions in (\ref{equationforzp}) and (\ref{aprime}) with $b=0$
and $\psi=\frac{\pi}{4}$,
\begin{align}
\label{equationforzp1}
&z=\int_0^r d\tilde r\frac{ f^2 }{\sqrt{(2f^2+1)  \tilde r^4 +q^2}}
\\
&a=\int_0^r d\tilde r\frac{-q }{\sqrt{(2f^2+1) \tilde  r^4 +q^2}} \label{aprime1}
\end{align}
Note that the range of these functions is finite.

The chemical potential is given by difference in the value of the gauge field at $r=\infty$ and $r=0$ (where
we remember the normalization in (\ref{magneticfields}))
\begin{align}
\mu= \frac{\sqrt{\lambda}}{2\pi}[a(\infty)-a(0)]=-\frac{\sqrt{\lambda}}{2\pi}\frac{|q|^{\frac{1}{2}}}{4(2f^2+1)^{\frac{1}{4}}}
{\mathcal B}[\frac{1}{4},\frac{1}{4}]~{\rm sign}(q)
 \label{aprime2}
\end{align}
${\mathcal B}[s,t]=\Gamma[s]\Gamma[t]/\Gamma[s+t]$ is Euler's beta-function.
Now, remembering that the charge density is defined by $\rho= \frac{\sqrt{\lambda} N}{4\pi^3}q$, we can write
the expression for the charge density as a function of the chemical potential,
\begin{align}
\boxed{
\rho =-{\rm sign}(\mu)~ \frac{16\sqrt{2f^2+1}~N }{\pi{\mathcal B}[\frac{1}{4},\frac{1}{4}]^2\sqrt{\lambda}}\mu^2
}
\end{align}
For a generic value of $f^2$ in the range of interest, say $f^2=1/2$,
\begin{align}
\rho
\sim -{\rm sign}(\mu)~\frac{.13}{\sqrt{\lambda}} N\mu^2  ~~{\rm with}~f^2=1/2
\end{align}
   One might compare this with the free field theory where
\begin{align}
   \rho_0=-\frac{N}{(2\pi)^2}\int d^2k \theta(\mu-|k|) = -\frac{N}{4\pi}~\mu^2\sim- .080 N\mu^2
\end{align}

The scaling with $\mu^2$ is a consequence of conformal invariance.
In a theory where the U(1) current is conserved, the Debye screening mass (the inverse of the Debye
screening length) can be derived from the charge density-chemical
potential relationship by taking a derivative of the charge density by the chemical potential\cite{fradkin},

\begin{align}\boxed{
L_D^{-1}=\frac{d}{d\mu}\rho= \frac{32\sqrt{2f^2+1}~N }{\pi{\mathcal
    B}[\frac{1}{4},\frac{1}{4}]^2\sqrt{\lambda}}|\mu|
=\frac{8}{\sqrt{\pi}}\frac{ (1+2f^2)^{\frac{1}{4}} }{
  B[\frac{1}{4},\frac{1}{4}] }\frac{ \sqrt{N} }{ \lambda^{\frac{1}{4}}
} |\rho|^{\frac{1}{2}} }\end{align}

It is interesting that coefficients of $N\mu^2$ in the strongly
coupled theory and the free field theory and the Debye screening
lengths differ by a factor of $\frac{1}{\sqrt{\lambda}}$ which can be
significant in the large $\lambda$ limit.  The Debye screening length
as a function of gate voltage is a quantity which could in principle
be measured in a relativistic condensed matter system such as
graphene. The coupling constant could also be varied, in principle, by
changing the dielectric constant of the vicinity of the material.
Such a measurement would be an interesting test of the hypothesis that
the relativistic material could be in a strongly correlated state that
is described by a conformal field theory similar to the one which
solves the D3-D7 model.

\section{Conformal field theory in a magnetic field}

The equation for the angle $\psi$ with $b\neq 0$ and  $q=0$  is
\begin{align}
\frac{\ddot\psi}{1+\dot\psi^2}+\dot\psi\left[ 1+\frac{2r^4[(f^2+4\cos^4\psi)(f^2+4\sin^4\psi) -f^4]
}{(f^2+4\cos^4\psi)(f^2+4\sin^4\psi) (b^2+r^4)-f^4r^4+(2bc)^2}\right] + \nonumber \\
+\frac{4\sin2\psi\left[[f^2 (   b^2+r^4)-r^4\sin^22\psi] \cos2\psi   -2b^2(\psi-\pi/4)\sin2\psi \right]
}{(f^2+4\cos^4\psi)(f^2+4\sin^4\psi) (b^2+r^4)-f^4r^4+(2bc)^2}=0
\label{equationforpsiwithq=0}\end{align}
This equation is solved by $\psi=\tfrac{\pi}{4}$.  The magnetization as a function of field
is given by the derivative of the vacuum energy by the field,
\begin{align}
M =- \frac{\partial}{\partial B}F = -\frac{1}{V_{2+1}}\frac{\partial}{\partial B}{\mathcal R}
\end{align}
where we have recognized that the energy as a function of field is
simply the negative on-shell Routhian, which we can find by setting
$q=0$, $\psi'=0$ in (\ref{routhian1}).  We obtain
\begin{align}
{\mathcal R}={\mathcal N} \int_0^\infty dr \sqrt{(f^2+1)^2
  (b^2+r^4)-f^4r^4}
\label{routhianmag}
\end{align}
The above integral diverges.  It can be defined by subtracting the
$b=0$ energy from it.  Then the result is finite.  Then, the
derivative by the field renders the remaining integral finite,
\begin{align}
&~~M=- \frac{\lambda
    N}{8\pi^4}\left(\frac{2\pi}{\sqrt{\lambda}}\right)^{\frac{3}{2}}\sqrt{B}
  \int_0^\infty d r \frac{(f^2+1)^2}{\sqrt{(f^2+1)^2
      (1+r^4)-f^4r^4}}~{\rm sign}(B)\nonumber \\ &\boxed{ M=-\frac{
      (f^2+1)^{\frac{3}{2}}\lambda^{\frac{1}{4}} N}{
      2(2\pi)^{\frac{5}{2}}(2f^2 +1)^{\frac{1}{4}} } {\mathcal
      B}[\tfrac{1}{4},\tfrac{1}{4}] \sqrt{B}~{\rm sign}(B) } \nonumber
  \\ &~~M\sim -(0.06)\lambda^{\frac{1}{4}}N\sqrt{B}~{\rm
    sign}(B)~~({\rm with}~f^2=1/2)
\label{magnetization}
\end{align}
The sign, $M\sim -{\rm sign}(B)$ is a result of diamagnetism, which is
what is expected for electrons in a magnetic field.  We can compare
this with the result for free Fermions.  There, all negative energy
states are occupied and their spectrum in a magnetic field is that of
relativistic Landau levels, given by $E_n=-\sqrt{2|B|n}$, with
$n=0,1,2,...$. The degeneracy of each Landau level is
$\frac{|B|}{2\pi}$.  The ground state energy is
$$E_0=-N\frac{|B|}{2\pi}\sum_1^\infty \sqrt{2|B|n}=-
\frac{|B|^{\frac{3}{2}}N}{\sqrt{2}\pi}\zeta(-1/2)$$ where we have
defined the infinite sum by zeta-function regularization and the value
of the zeta-function is $\zeta(-1/2)=-0.21$.  The vacuum energy is
positive and
\begin{align}
M=- \frac{3}{\sqrt{8\pi^2}}(0.21)N\sqrt{|B|}~{\rm sign}(B)\sim -0.07 N\sqrt{B}~{\rm sign}(B)
\end{align}
It is again interesting that the diamagnetism of strongly and weakly
interacting Fermions differ by a factor of $\lambda^{\frac{1}{4}}$.

Now, we observe that, with a finite magnetic field, at $r\to0$, the
angle $\psi(0)$ must solve the equation
\begin{equation}
f^2     \cot2\psi(0)   -2 (\psi(0)-\pi/4)  =0
\end{equation}
The only solution of this equation is $\psi(0)=\frac{\pi}{4}$, which
fixes the value that $\psi(r)$ must take at $r\to0$.  However, if we
study the linearized equation in this region, we find that, unlike the
case that we studied in the previous sections where the constant
$\psi=\frac{\pi}{4}$ was an isolated solution, here, it is not
isolated.  There are solutions close by where $\psi(r)$ depends on $r$
and approaches $\frac{\pi}{4}$ when $r$ goes to zero and infinity.  To
see this, the linearized equation in the small $r$ region is
\begin{align}
\ddot\psi+\dot\psi
-\frac{8(\psi-\pi/4)
}{(1+f^2) }=0
\label{linearizedequationforpsiwithq=0}\end{align}
and it has solutions
\begin{align}\label{smallrasymptotic}
\psi(r)=\frac{\pi}{4}+\kappa_1r^{\frac{1}{2}+ \frac{1}{2}\sqrt{1+\frac{32}{1+f^2}}}
+\kappa_2r^{\frac{1}{2}- \frac{1}{2}\sqrt{1+\frac{32}{1+f^2}}}+\ldots
\end{align}
We see that the solution (\ref{smallrasymptotic}) converges to
$\frac{\pi}{4}$ if we choose the constant $\kappa_2=0$. This requires
tuning the asymptotic behavior of the solution at large $r$, which we
shall examine in more detail in the following.  In this way, we see
that, when there is a magnetic field present, there are solutions of
the equation for $\psi(r)$ which are infinitesimally close to the
constant $\psi=\frac{\pi}{4}$ for all values of $r$.  This is in
contrast to the case without a magnetic field where any infinitesimal
deviation of $\psi(r)$ from $\frac{\pi}{4}$ at large $r$ led to a
non-infinitesimal deviation at small $r$.  In the present case with
magnetic field, we can study solutions which deviate from
$\frac{\pi}{4}$ by a small amount perturbatively. We shall spend the
remainder of this section studying the properties of these solutions.

Before we begin, we note that, as soon as $\psi(r)$ deviates from the
constant $\frac{\pi}{4}$, the solution violates all three of the
discrete symmetries, parity, charge conjugation and CP. That it
violates C and CP can be seen from the equation (\ref{aprime}) which
requires $a(r)$ to be non-zero, at least for intermediate values of
$r$.  However, we are still free to set the parameter $q$, and
therefore the charge density to zero.  In the next Section, we will
examine what happens when we turn on a $q$ of infinitesimal
magnitude.

In the appendix, we have studied (\ref{equationforpsiwithq=0}) perturbatively. Parameterizing the asymptotic behavior of $\psi(r)$ by two parameters $m$ and $c$, $\psi(r\to\infty)=\frac{\pi}{4}+\frac{m}{r^{\Delta_-}}+\frac{c}{r^{\Delta_+}}+\ldots$, we found that they are related to each other in the following way

\begin{align}
c = \left\{ \begin{array}{ll}
         {-2^{u+\nu}\left(\frac{1+f^2}{\sqrt{1+2f^2}}b\right)^{2\nu+1}
{ \frac{{\Gamma[1-u+\nu]~\Gamma[1+\nu]}~\Gamma\left[{u+v+1\over2}\right]
\Gamma\left[{u+v+2\over2}\right]}{{\Gamma\left[1+2\nu\right]}~\Gamma[u]~\Gamma[1-u]~\Gamma\left[\nu+{3\over2}\right]}}}m & \mbox{if $\nu > 0$};\\
        (\frac{1+f^2}{\sqrt{1+2f^2}})^{2\nu+1}{\left[2^{\nu}\frac{\Gamma[1-2\nu]}{\Gamma[1-\nu]\Gamma[1-\nu-u]}\frac{(v+u)}{2\nu+1}-2^u \frac{\Gamma\left[{u+v+1\over2}\right]\Gamma\left[{u+v+2\over2}\right]}{\Gamma[u]\Gamma[1-u]\Gamma\left[\nu+{3\over2}\right]}\right]
\over\frac{\Gamma[1+2\nu]}{\Gamma[1-u+\nu]\Gamma[1+\nu]}(\frac{1}{2})^{\nu}}m & \mbox{if $\nu < 0$}.\end{array} \right.
\label{result}\end{align}

where $\nu$ and $u$ are given by
\begin{align}
\nu=-\frac{1}{2}+\sqrt{\frac{50f^2-23}{16(1+2f^2)}} ~~,~~
u=\frac{\sqrt{33+f^2}}{4\sqrt{1+f^2}}
\label{definition}
\end{align}

\begin{figure}[ht]
\begin{center}
\includegraphics[scale=0.4]{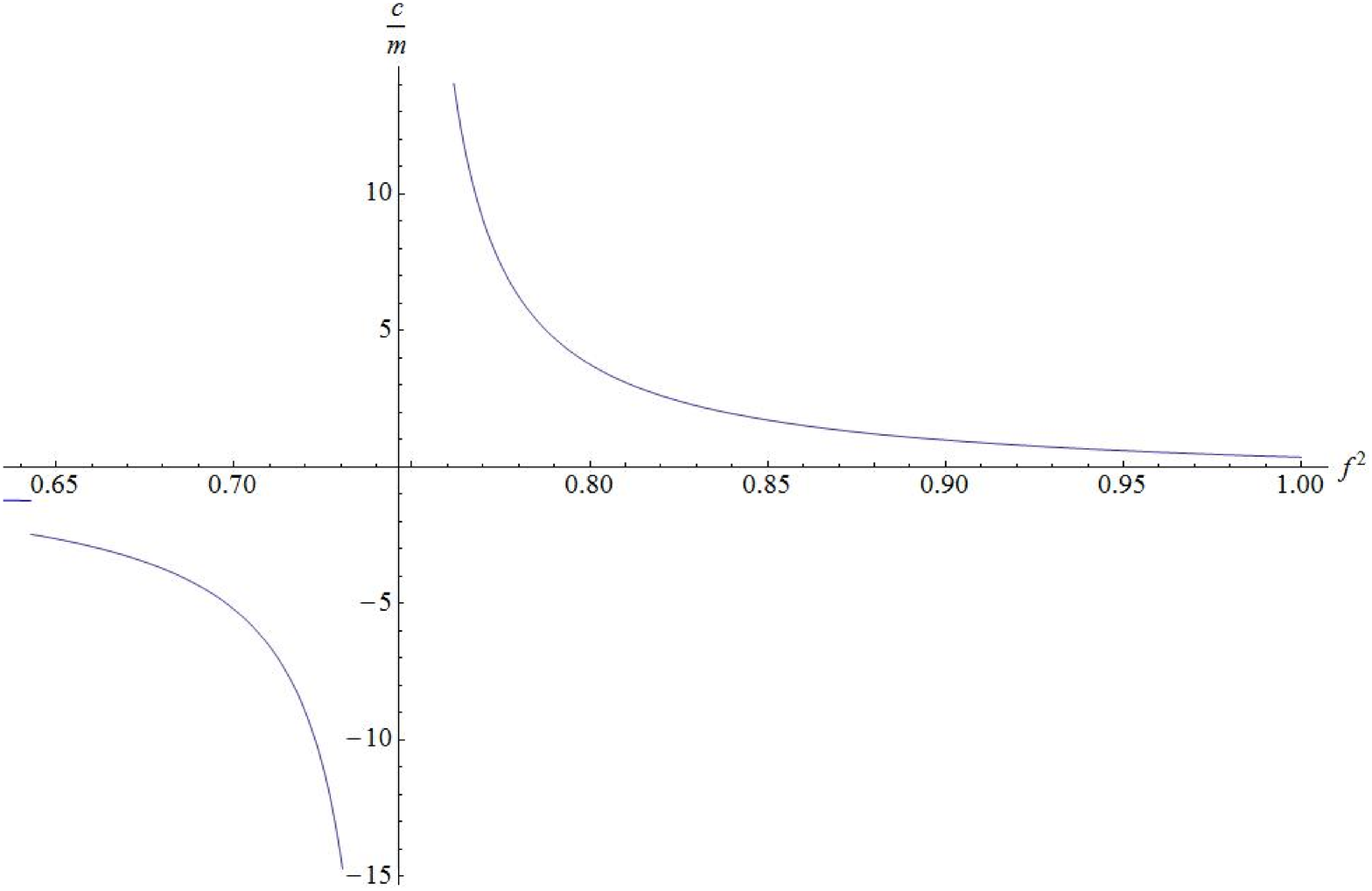}
\end{center}
\caption{\label{fig-throat} {\footnotesize The ratio of condensate to
    mass parameter $\frac{c}{m}$ is plotted versus the internal flux
    parameter $f^2$.  The plot exhibits a discontinuity at $\nu=0$ and
    a singularity at $1-u+\nu=0$}, the value where the condensate
  changes sign, and also where spontaneous breaking of chiral symmetry
  can take place.}
\end{figure}

In equation (\ref{result}) we see that the function $c(m,f^2)$ changes in character as $f^2$ is varied from below
to above the value $f^2=\frac{9}{14}$ at which $\nu=1$.  In fact, as depicted in figure \ref{fig-throat}, the function is discontinuous there.
The behavior resembles a quantum phase transition - for a fixed value of $m$, $c$ jumps in a discontinuous way
as $f^2$ is varied.
We note that this special value   $f^2=\frac{9}{14}$ is just the value where
the exponents $\Delta_+$ and $\Delta_-$,
which are interpreted as scaling dimensions of the chiral condensate and mass parameter, both obey the lower bound on
dimensions that is required in a unitary conformal field theory, they have values $(\Delta_+,\Delta_-)=(\frac{5}{2},\frac{1}{2})$.
At that bound, an alternative quantization sets in.
When $f^2<\frac{9}{14}$, there is an alternative quantization where the interpretation of the mass and chiral condensate
are interchanged, whereas when $f^2>\frac{9}{14}$, the quantization that we describe is unique.

We note one other interesting behavior which occurs as $f$ is varied.
When it achieves a value $f^2\sim.74$, so that $1-u+\nu=0$, the ratio
$\frac{c}{m}$ in (\ref{result}) seems to diverge. This behavior is
depicted in figure \ref{fig-throat}. We interpret it as the
condensate $c$ remaining non-zero as $m$ is put to zero, that is,
spontaneous symmetry breaking, the presence of a nonzero condensate
when the source is put to zero.  In addition, at this point, the sign
of the condensate versus the sign of the magnetic field flips, for
$f^2>.74$ the sign of $c$ and the sign of $m$ are identical, for
$f^2<.74$ the signs are opposite.\\

Here we have assumed that $m$ and $c$ are small parameters (in magnetic field units).
This has allowed us to linearize and solve the equation for $\psi(r)$ for all values of r.
The existence of a value of $f^2$ at which $m=0$ but $c$ may differ from zero is a
result of this analysis.  We comment here that the result is possibly more robust and could
well persist beyond our linear approximation. This follows from the fact that, if we fix $c$ and
solve the theory to extract $m(c,f^2)$ for differing values of $f^2$, we find that $m$ generically varies
smoothly and changes sign as $f^2$ is swept across the domain $\frac{23}{50}<f^2<1$.  This
means that $m(c,f^2)$ has a zero at some value of $f^2$.  In the linearized limit, $m(c,f^2)$ becomes
a linear function of $c$ and we see that the zero occurs at $f^2\sim 0.74$

We have found that the conformal dimensions of the operator which is
dual to fluctuations of the angle $\psi$ lie in an interesting range
when $\frac{23}{50}\leq f^2\leq 1$. A summary of some special values
of $f^2$ in this range are given in Table \ref{specialf}.

\begin{table}
 \centering
\caption{ Special values of $f^2$}
 \begin{tabular}{ l | l | l }
    \noalign{\smallskip}\hline
    $f^2$ & $\Delta_+$,$\Delta_-$ & ~~ \\ \noalign{\smallskip}\hline\noalign{\smallskip}
    $\frac{23}{50}$ & $\frac{3}{2},\frac{3}{2}$ &{\rm stability~bound} \\ \noalign{\smallskip}\hline\noalign{\smallskip}
    $\frac{1}{2}$ & $2,1$ & {\rm classical~dimensions} \\ \noalign{\smallskip}\hline\noalign{\smallskip}
   $\frac{9}{14}$ & $\frac{5}{2},\frac{1}{2}$ & {\rm unitarity~bound} \\ \noalign{\smallskip}\hline  \noalign{\smallskip}
   $0.74$ & $m=0,c\neq$0 & {\rm chiral~symmetry~breaking} \\ \noalign{\smallskip}\hline\noalign{\smallskip}
   $1$ & $3,0$ & {\rm marginal} \\
    \noalign{\smallskip}\hline
  \end{tabular}
\caption{\label{specialf} \footnotesize Table of special values of $f^2$. For
  stability of the system to fluctuations, it is necessary that
  $f^2>\frac{23}{50}$.  When $f^2<\frac{9}{14}$, both dimensions
  satisfy a unitarity bound and there exists an alternative
  quantization where the mass and the condensate can be interchanged.
  At $f^2\approx .74$, magnetic catalysis of chiral symmetry breaking
  is can take place.  When $f^2=1$, the mass operator is exactly
  marginal.}
\end{table}

\section{Finite density in presence of magnetic field}

As we have remarked in the previous sections, the equation of motion
(\ref{equationforpsi}) has the constant solution $\psi=\pi/4$ only
when either $b=0$, $q=0$ or both. When both $b$ and $q$ are nonzero,
there is no constant solution.  When $q$ and $b$ were set to zero, we
saw that the constant solution $\psi=\frac{\pi}{4}$ was isolated.  As
soon as we turn on a non-constant behavior of $\psi(r)$ at $r\to
\infty$, the solution near $r\to 0$ deviates by a large amount from
$\frac{\pi}{4}$.  Then, in the previous section, we found that when
$b$ is nonzero, but we still tune $q$ to zero, the constant solution
is no longer isolated and we explicitly found a solution which differs
from the constant by a small amount for all values of $r$.  In this
Section, we shall examine what happens when we turn on a small value
of $q$.

In the small $r$ region, the non-constant solution must go to a zero of the last term in (\ref{equationforpsi}).  That is,
it must solve the equation
\begin{equation}
\label{origin}
f^2     \cot2\psi(0) + \frac{q}{b}  -2 (\psi(0)-\pi/4)  =0
\end{equation}
When $\frac{q}{b}=0$, this equation is solved by
$\psi(0)=\frac{\pi}{4}$, and the constant solution is allowed and, as
we saw in the previous section, even when it is not constant,
$\psi(r)$ approached $\frac{\pi}{4}$ when $r$ approaches either 0 or
$\infty$.  Now, when $\frac{q}{b}$ is nonzero, this is no longer the
case, there is no constant solution and the non-constant must approach
a zero of (\ref{origin}) which is no longer $\frac{\pi}{4}$.  If we
consider the case where $\frac{q}{b}$ is small, it is solved by
\begin{equation}
\phi(0)= \frac{\frac{q}{b}}{2(1+f^2)}
\end{equation}
It is consistent that the deviation of $\psi(r)$ from $\frac{\pi}{4}$
is of order $\frac{q}{b}$.  If we consider
\begin{equation}
\psi(r)=\frac{\pi}{4}+\phi(r)
\end{equation}
and Taylor expand (\ref{equationforpsi}) to first order in $\phi$ and
$q$, we obtain equation (\ref{eqmo}), which we recopy here for the
reader's convenience:
\begin{align}
 [b^2(1+f^2)^2+r^4(1+2f^2)]r^2\phi''
+2[b^2(1+f^2)^2+2r^4(1+2f^2)]r\phi'+
 \nonumber \\ +8[r^4(1-f^2)-b^2(1+f^2)]\phi
=-4bq
\label{eqm2}
\end{align}
The solution of the (\ref{eqm2}) is given by
\begin{equation}
\phi=\phi_H+\int_{0}^{\infty} dy G(x-y)f(y)
\end{equation}
where $\phi_H$ is the general homogenous solution and $G(x-y)$ is the
Green function. The Green function is given by
\begin{equation}
G(x,y)=g(x)h(y)\Theta(y-x)+g(y)h(x)\Theta(x-y)
\end{equation}
where $g(x)$ is the regular solution at origin, $h(x)$ is the
regular solution at the boundary and $\Theta$ is the step
function. Having only one regular solution at the origin, we are
forced to choose the linear combination of $P$ and $Q$ that we found
in (\ref{smallrsolution}) . Both of $P$ and $Q$ are regular at the
boundary and vanish. We specify the solution by specifying the
boundary condition at $r=\infty$ so that the Green function part does
not contribute to the coefficient of term that dies slower, $m$; so,
we choose $h(x)$ to be $Q(x)$. We are just left to find the right
normalization of our Green function so that

\begin{equation}
 (1-x^2)G(x,y)''-2xG(x,y)'+\left(\nu(\nu+1)-\frac{u^2}{1-x^2}\right)G(x,y)=\delta(x-y)
\end{equation}
where the primes are differentiation with respect to $x$. This equation can be written as

\begin{equation}\label{sturm}
 {d\over dx}((1-x^2)G(x,y)')+\left(\nu(\nu+1)-\frac{u^2}{1-x^2}\right)G(x,y)=\delta(x-y)
\end{equation}
Integrating the above equation, we get the following constraint
\begin{align}
(1-x^2)G(x,y)'|_{y+\epsilon}-(1-x^2)G(x,y)'|_{y-\epsilon}=1
\end{align}
which can be written as
\begin{equation}
\label{const}
(1-x^2)W[g(x),h(x)]=1
\end{equation}
where $W[h(x),g(x)]$ is the Wronskian.  The (\ref{sturm}) has a
form of Sturm-Liouville equation. It is a well-known property of such
an equation that the Wronskian of the solutions is given by
 \begin{equation}
W[h(x),g(x)]\propto\frac{1}{(1-x^2)}
 \end{equation}
 which ensures that we can satisfy the constraint in (\ref{const}). We
 just need to evaluate the Wronskian at a given point to fix the
 normalization.  Substituting the asymptotic expansions of $P$ and
 $Q$, we get
 \begin{align}
 \label{wrons}
 W[Q,P]=(\frac{1}{2})^{2\nu+1}(2\nu+1)~\frac{{\Gamma(1+2\nu)}
 }{\Gamma(1-u+\nu)\Gamma(1+\nu)}
 \frac{\Gamma(\nu+u+1)}{\Gamma(\nu+3/2)}{\sqrt{\pi}\over (1-x^2)}
\end{align}
Now we can use (\ref{wrons}) to find the Wronskian. Using the linear
combination (\ref{smallrsolution}) as $g(x)$, we find that the
Wronskian is given by

\begin{align}
W[Q,g]&=(\frac{1}{2})^{2\nu+1}~\frac{{\Gamma(2+2\nu)}
}{\Gamma(1-u+\nu)\Gamma(1+\nu)}~
\frac{\Gamma(\nu+u+1)}{\Gamma(\nu+3/2)}{\sqrt{\pi}\over
  (1-x^2)}\nonumber \\ &=\frac{1}{{\mathcal{C}}}{1\over (1-x^2)}
\end{align}
So, we need to define $h(x)$ in the following way to get the right
normalization
\begin{equation}
h(x)=\mathcal{C} Q^u_{\nu}(x)
\end{equation}
Putting all the parts together, we can finally write the complete
expression of $\psi$

\begin{align}
\psi={\pi\over4}+ \phi_H -
\frac{q}{b}\frac{\mathcal{C}}{\sqrt{rb^{-\frac{1}{2}}}}
Q^u_{\nu}(x)\int_0^{\infty} dy \left(P(\nu,u,y)+\frac{c_2}{c_1}~
Q(\nu,u,y)\right) \frac{1}{(1-y^2)(1+f^2)^{3\over4}}\left({
  \frac{y^2-1}{1+2f^2}} \right)^{\frac{1}{8}}
\end{align}
where $\phi_H$ is the solution to homogeneous equation and the rest
are as listed below:
\begin{equation}
x=\sqrt{1+\frac{1+2f^2}{(1+f^2)^2}\frac{r^4}{b^2}}
\end{equation}
\begin{equation}
\frac{c_2}{c_1}=- \frac{2^{\nu+u+1}}{\sqrt{\pi}}
\frac{\Gamma\left[\frac{u+\nu+1}{2}\right]\Gamma\left[\frac{u+\nu+2}{2}\right]}{\Gamma[u]\Gamma[\nu+u+1]\Gamma[1-u]}
\end{equation}
\begin{equation}
{1\over \mathcal{C}}=(\frac{1}{2})^{2\nu+1}~\frac{{\Gamma(2+2\nu)}
}{\Gamma(1-u+\nu)\Gamma(1+\nu)}
\frac{\sqrt{\pi}\Gamma(\nu+u+1)}{\Gamma(\nu+3/2)}
\end{equation}

In this section, we have derived an explicit solution of equation of
motion in presence of magnetic field and finite density. This solution
shows finite deviation from constant solution.

\section{Conclusions}

In this Paper, we have examined a regime where a mass operator,
magnetic field and charge density can be turned on and the
modification of the D3-D7 conformal field theory can be studied
perturbatively.  In Sections 5 and 6 we solved the leading
perturbation explicitly.  For this solution, we can compute the
relationship between the coefficient of the mass operator and the
chiral condensate for interesting values of $f^2$. This function is
depicted in figure \ref{fig-throat}.  It has the interesting feature
that, at $f^2=.74$, the relationship becomes singular.  We interpret
the singularity as a signal of spontaneous symmetry breaking.

Moreover, as a byproduct, we have found holographic expressions for
the Debye screening length as a function of fermion density and of the
diamagnetic moment of the ground state as a function of magnetic
field.  They both exhibit a dependence on the charge density and
magnetic field consistent with scale symmetry.  They also have a mild
dependence on the coupling constant which could in principle be
interesting to compare with a real material, such as graphene where
these quantities can be measured.  It would be interesting to check
the functional form in order to test the hypothesis that graphene is
scale invariant.  It would also be interesting to see whether these
quantities in graphene differ in any significant way from the free
field expressions, and whether the deviation is in the same direction
as predicted by the strong coupling formula.

\appendix
\section{Appendix}
We begin by considering
a solution of equation (\ref{equationforpsiwithq=0}) which deviates from the constant by a small amount,
\begin{equation}
\psi(r)=\frac{\pi}{4}+\phi(r)
\end{equation}
and Taylor expand (\ref{equationforpsi}) to first order in $\phi$ and
$q$, we obtain
\begin{align}
 [b^2(1+f^2)^2+r^4(1+2f^2)]r^2\phi''
+2[b^2(1+f^2)^2+2r^4(1+2f^2)]r\phi'+
 \nonumber \\ +8[r^4(1-f^2)-b^2(1+f^2)]\phi
=-4bq
\label{eqmo}
\end{align}
which we shall now try to solve over the full range of $r$.  Note that
(\ref{eqmo}) reduces to our small $r$ linearization in
(\ref{smallrasymptotic}) at $r\to0$.  The transformation
\begin{align}
y(x)=\sqrt{\frac{r}{b^{\frac{1}{2}}}}\phi(r)
~~,~~
x=\sqrt{1+\frac{1+2f^2}{(1+f^2)^2}\frac{r^4}{b^2}}
\end{align}
puts (\ref{eqmo}) in the following form, which in the case of $q=0$ is
the standard form of the associated Legendre equation
 \begin{equation}
 (1-x^2)y''-2xy'+\left(\nu(\nu+1)-\frac{u^2}{1-x^2}\right)y=-\frac{q}{b}\frac{1}{(1-x^2)(1+f^2)^{3\over4}}\left({
     \frac{x^2-1}{1+2f^2}} \right)^{\frac{1}{8}}
 \label{eqmotion}
 \end{equation}
where the parameters are
\begin{align}
\nu=-\frac{1}{2}+\sqrt{\frac{50f^2-23}{16(1+2f^2)}} ~~,~~
u=\frac{\sqrt{33+f^2}}{4\sqrt{1+f^2}}
\end{align}
For future reference, we note that $u$ is always positive but $\nu$
can change sign.  One can write $\nu$ in terms of $\Delta_\pm$ as
\begin{align}
\Delta_\pm=\frac{3}{2}\pm\sqrt{\frac{9}{4}-8\frac{1-f^2}{2f^2+1}}
 =\frac{3}{2}\pm\sqrt{\frac{50f^2-23}{4(1+2f^2)}}
\end{align}
\begin{align}
\nu=-\frac{5}{4}+\frac{\Delta_+}{2} ~~,~~\nu=\frac{1}{4}-\frac{\Delta_-}{2}
\end{align}
The Legendre equation has the general solution
\begin{equation}
\phi=\frac{1}{\sqrt{r}}\left(c_1P\left[\nu,u,\sqrt{1+{(1+2f^2)\over(1+f^2)^2}\frac{r^4}{b^2}}\right]+c_2
Q\left[\nu,u,\sqrt{1+{(1+2f^2)\over(1+f^2)^2}\frac{r^4}{b^2}}\right]\right)
\label{solutionofeqmo}
\end{equation}
The associated Legendre functions are given by Hypergeometric
functions as below
\begin{equation}
\label{P}
P^u_{\nu}(x)=\frac{1}{\Gamma[1-u]}\left[
  \frac{1+x}{-1+x}\right]^{\frac{u}{2}} {_2F_1}\left[-\nu, \nu+1;
  1-u;\frac{1-x}{2}\right]
\end{equation}
\begin{equation}
Q^u_{\nu}(x)=
\frac{\sqrt{\pi}}{2^{\nu+1}}\frac{\Gamma[\nu+u+1]}{\Gamma[\nu+3/2]}
\frac{(-1+x^2)^{u/2}}{x^{\nu+u+1}}
{_2F_1}\left[\frac{u+\nu+1}{2},
\frac{u+\nu+2}{2}; \nu+3/2;\frac{1}{x^2}\right]
\label{Q}\end{equation}
where we have defined phases so that the functions are real for
$x>1$.  Both of the solutions (\ref{P}) and (\ref{Q}) diverge at
$r\to0$ (which corresponds to $x\to 1^+$) and converge to zero at
$r\to\infty$ (which corresponds to $x\to\infty$).

To find a solution which is finite at $r\to0$, we need to adjust the
two constants, $c_1$ and $c_2$ in (\ref{solutionofeqmo}) so that the
singularity cancels.  Since $r\to0$ corresponds to $x\to1^+$, we shall
need to study $_2F_1[a,b,c,z]$ in the regions where $z=0$ for $P$ and
$z=1$ for $Q$, respectively.  It is easy to study $P$ near $x\to 1^+$.
From the definition of Hypergeometric functions,
\begin{equation}
{_2F_1[a,b,c,z]}=\sum_{n=0}^{n=\infty}\frac{(a)_n(b)_n}{(c)_n}\frac{z^n}{n!}
~~,~~
(a)_n=\frac{\Gamma[a+n]}{\Gamma[a]}
\end{equation}
we see that the leading term in the series is normalized to one,
\begin{equation}
{_2F_1[a,b,c,0]}=1
\end{equation}
Then, from (\ref{P}), we see that the asymptotic behavior of $P$ is
given by
\begin{equation}
P^u_{\nu}(x)\simeq \frac{1}{\Gamma[1-u]}\left[
  \frac{2}{-1+x}\right]^{\frac{u}{2}}~~,~~x\simeq 1^+
\end{equation}

Now, let us consider $Q$. We note that, for the values of $u$ and
$\nu$ of interest, $Q$ is divergent at $z=1$.  Since the asymptotic
small $r$ behavior of the solutions of the Legendre equations exhibit
only one divergent solution, we know from the outset that the
divergent nature of $Q$ must be similar to that of $P$.

Euler's formula is an integral representation of $F$ which is valid
when $\Re(c)>0$ and $\Re(b)>0$,
\begin{equation}\label{integralrepresentation}
{_2F_1[a,b,c,z]}=\frac{\Gamma[c]}{\Gamma[c-b]}{\Gamma[b]}\int_{0}^{1}t^{b-1}(t-1)^{c-b-1}(1-tz)^{-a}dt
\end{equation}
Using it, one can show that
\begin{equation}
{_2F_1[a,b,c,z]}=(1-z)^{c-a-b}{_2F_1[c-a,c-b,c,z]}
\label{ftransformation}
\end{equation}
Using the same representation one can show that if $\Re(c-a-b)>0$
\begin{equation}
{_2F_1[a,b,c,1]}=\frac{\Gamma[c]\Gamma[c-a-b]}{\Gamma[c-a]\Gamma[c-b]}
\label{fidentity1}
\end{equation}
for $\Re(b)<0$ the above result can be generalized using analytical
continuation.

In our case of interest,  $Q$ in Eq.~(\ref{Q}) has
\begin{align}
& b=\frac{u+\nu+2}{2}>0\\ &
c-a-b=-u<0
\end{align}
which is not in the domain of applicability of (\ref{fidentity1}).  If
we use the transformation (\ref{ftransformation}), we can change the
arguments of $F$ so that (\ref{fidentity1}) can be applied.  Then we
get
\begin{equation}
{_2F_1[a,b,c,z]}=(1-z)^{-u}{_2F_1[c-a,c-b,c,z]}
\end{equation}
where, now
\begin{align}
c-(c-b)-(c-a)&=a+b-c \nonumber \\ &=u>0
\end{align}
The result is
\begin{equation}
{_2F_1[a,b,c,z]}\simeq(1-z)^{-u}\frac{\Gamma[c]\Gamma[a+b-c]}{\Gamma[a]\Gamma[b]}
\label{resultid}\end{equation}
and we find the asymptotic behavior of $Q$,
\begin{equation}
Q^u_{\nu}(x)\simeq\frac{\sqrt{\pi}\Gamma[\nu+u+1]}{2^{\nu+1}}
\left(\frac{1}{2(-1+x)}\right)^{u/2}\frac{\Gamma[u]}
{\Gamma[\frac{u+\nu+1}{2}]\Gamma[\frac{u+\nu+2}{2}]}~~,~~x\simeq
1^+
\end{equation}
which shows the expected divergent behavior.  Then the regular
solution turns out to be (\ref{solutionofeqmo}) (which we recopy here)

\begin{equation}
\phi=\frac{1}{\sqrt{r}}\left(c_1P\left[\nu,u,\sqrt{1+{1+2f^2\over(1+f^2)^2}\frac{r^4}{b^2}}\right]
+c_2~
Q\left[\nu,u,\sqrt{1+{1+2f^2\over(1+f^2)^2}\frac{r^4}{b^2}}\right]\right)
\nonumber
\end{equation}
where $c_1$ and $c_2$ are related by
\begin{equation}
\frac{c_2}{c_1}=- \frac{2^{\nu+u+1}}{\sqrt{\pi}}
\frac{\Gamma\left[\frac{u+\nu+1}{2}\right]\Gamma\left[\frac{u+\nu+2}{2}\right]}{\Gamma[u]\Gamma[\nu+u+1]\Gamma[1-u]}
\label{smallrsolution}\end{equation}

We have now found an acceptable non-singular solution in the small
$r\to0$ regime.  It is given by equation (\ref{solutionofeqmo}) with
the constants related by (\ref{smallrsolution}).  We now need to find
what this implies for the large $r$ behavior of the solution.  In
particular, it should fix the relationship between the coefficients of
the two asymptotic behaviors of the solution at large $r$ which are
displayed in equation (\ref{fluctuations}), that is, it should fix the
relationship between $c$ and $m$.  For large $r$ one gets
\begin{equation}
x=1+\sqrt{{1+2f^2\over(1+f^2)^2}\frac{r^4}{b^2}}\simeq{\sqrt{1+2f^2}\over(1+f^2)}
\frac{r^2}{b}
\end{equation}
Using the same integral representation (\ref{integralrepresentation}),
one can show that
\begin{equation}
{_2F_1[a,b,c,z]}=(1-z)^{-a}~{_2F_1\left[a,c-b,c,\frac{z}{z-1}\right]}
\end{equation}
For $P$ in the regime where  $\frac{1}{x}\simeq 0$ we get
\begin{align}
P^u_{\nu}[x]&\simeq\frac{1}{\Gamma[1-u]}\left(1+\frac{u}{x}\right)
~{_2F_1}\left[-\nu, \nu+1; 1-u;\frac{1-x}{2}\right] \nonumber \\ &
\simeq
\frac{1}{\Gamma[1-u]}\left(\frac{x}{2}\right)^{\nu}\left(1+\frac{u}{x}\right)
\left(1+\frac{\nu}{x}\right)
~{_2F_1\left[-\nu,-u-\nu,1-u,1-\frac{2}{x}\right]}
\label{secondeq} \\ &
\simeq \frac{1}{\Gamma[1-u]}\left(\frac{x}{2}\right)^{\nu}\left(1+\frac{\nu+u}{x}\right)
~\left[{_2F_1[-\nu,-u-\nu,1-u,1]}-\frac{2}{x}\left.
\frac{d}{dx}~_2F_1[-\nu,-u-\nu,1-u,x]\right|_{x=1}\right]\nonumber \\
\label{taylorexp}\end{align}
 Because $1>\nu\geq -\frac{1}{2}$, then we find that $
 1-u-(-u-\nu)-(-\nu)>0$, which allows us to use (\ref{fidentity1}).
 Using the identity (\ref{fidentity1}), we find that
\begin{equation}
{_2F_1[-\nu,-u-\nu,1-u,1]}=\frac{\Gamma[1-u]\Gamma[1+2\nu]}{\Gamma[1-u+\nu]\Gamma[1+\nu]}
\end{equation}

Another identity that we shall use is the following identity
\begin{equation}
\frac{d}{dz}~{_2F_1[a,b,c,z]}=\frac{ab}{c}{_2F_1[a+1,b+1,c+1,z]}
\end{equation}
Using this identity, we see that the second term in (\ref{taylorexp}) can be written as
\begin{align}
\left.
\frac{d}{dx}~_2F_1[-\nu,-u-\nu,1-u,x]\right|_{x=1}=\left.\frac{\nu(\nu+u)}{1-u}{_2F_1[1-\nu,1-u-\nu,2-u,x]} \right|_{x=1}
\label{deriv}\end{align}
Now, we should be more careful because for $0>\nu\geq -\frac{1}{2}$, we can not use the integral representation (\ref{integralrepresentation}).
In this case $c-a-b=2\nu<0$ and the (\ref{fidentity1}) can not be used. We shall therefore  consider the two cases $\nu>0$ and $\nu<0$ separately:

For $\nu>0$, using (\ref{deriv}) and (\ref{fidentity1}), we find out that

\begin{align}
\left.\frac{d}{dx}~_2F_1[-\nu,-u-\nu,1-u,x]\right|_{x=1}
 =\frac{\nu(\nu+u)}{1-u}\frac{\Gamma[2-u]\Gamma[2\nu]}{\Gamma[1-u+\nu]\Gamma[1+\nu]}
 \end{align}
Then one has
\begin{equation}
P^u_{\nu}(x)\simeq\frac{\Gamma[1+2\nu]}{\Gamma[1-u+\nu]\Gamma[1+\nu]}\left(\frac{x}{2}\right)^{\nu}~\left({1}+
{\mathcal O}(\frac{1}{x^2})\right)
\end{equation}
where the terms that fall off as $x^{\nu-1}$ get canceled by each other (as one expects from the asymptotic solutions of  (\ref{eqmotion}) for large values of $x$).\\

To deal with the $\nu<0$ case we can not use the Taylor expansion directly, as the derivative of $F$ at $x=1$ diverges. From the asymptotic solutions of  (\ref{eqmotion}), we know that to the two highest orders,  $P$ can either go like $x^\nu$ or $x^{-1-\nu}$. The $x^\nu$ part is easy to find, we can just use the (\ref{secondeq}), and look at $x\to \infty$.  If we subtract the $x^\nu$ term from $P$, because  $x^{-1-\nu}$ vanishes at infinity, the result would vanish at infinity. Considering this fact, we find the following limit

\begin{align}
 \lim_{x \to +\infty}\frac{_2F_1[-\nu,-u-\nu,1-u,1-\frac{2}{x}]-{_2F_1[-\nu,-u-\nu,1-u,1]}}{x^{-2\nu-1}} =\nonumber \\=\lim_{x \to +\infty}\frac{\frac{d~_2F_1\left[-\nu,-u-\nu,1-u,1-\frac{2}{x}\right]}{dx}}{\frac{d~x^{-2\nu-1}}{dx}}
\end{align}

which can be used to find the coefficient of $x^{-\nu-1}$ term in $P$
(the extra power of $\nu$ is because of presence of $x^\nu$ in (\ref{taylorexp})).
Using (\ref{deriv}), (\ref{ftransformation})and (\ref{fidentity1}), we find that
\begin{align}
\frac{d}{dx}~_2F_1\left[-\nu,-u-\nu,1-u,1-\frac{2}{x}\right]\simeq\frac{\nu(\nu+u)}{1-u}
\frac{\Gamma[2-u]\Gamma[-2\nu]}{\Gamma[1-\nu]\Gamma[1-u-\nu]}\left(\frac{2}{x}\right)^{2\nu}\left(\frac{2}{x^2}\right)
\end{align}
where $\frac{2}{x^2}$ comes from the chain rule used to change the variable of differentiation.

So we find that for $\nu<0$, $_2F_1\left[-\nu,-u-\nu,1-u,1-\frac{2}{x}\right]$ can be written as
\begin{align}
_2F_1\left[-\nu,-u-\nu,1-u,1-\frac{2}{x}\right]\simeq~_2F_1[-\nu,-u-\nu,1-u,1]-\nonumber \\-\frac{\nu(\nu+u)}{(1-u)(2\nu+1)}\frac{\Gamma[2-u]\Gamma[-2\nu]}{\Gamma[1-\nu]\Gamma[1-u-\nu]}\left(\frac{2}{x}\right)^{2\nu+1}
\end{align}

Plugging in the above results in (\ref{secondeq}), we find the following asymptotic expansion of $P$
\begin{equation}
P^u_{\nu}(x)\simeq \frac{\Gamma[1+2\nu]}{\Gamma[1-u+\nu]\Gamma[1+\nu]}\left(\frac{x}{2}\right)^{\nu}
\left(1+\frac{\nu+u}{x}\right)+\left(\frac{2}{x}\right)^{\nu+1}\nu(\nu+u)\frac{\Gamma[-2\nu-1]}{\Gamma[1-\nu]\Gamma[1-u-\nu]}
\end{equation}
where we have used the fact that $\Gamma[x+1]=x\Gamma[x]$. \\
We neglect the $\frac{u+\nu}{x}$ term, as it belongs to next orders. We then  have the following simplified equation of $P$ for $\nu<0$
\begin{equation}
P^u_{\nu}(x)\simeq \frac{\Gamma[1+2\nu]}{\Gamma[1-u+\nu]\Gamma[1+\nu]}\left(\frac{x}{2}\right)^{\nu}
+\left(\frac{2}{x}\right)^{\nu+1}\nu(\nu+u)\frac{\Gamma[-2\nu-1]}{\Gamma[1-\nu]\Gamma[1-u-\nu]}
\end{equation}
Around $\nu=-\frac{1}{2}$ the normalizable/un-normalizable nature of the two terms switches. Later, we will explore the behaviour of the solution around $\nu=-\frac{1}{2}$, and see if there is anything special about this point.
\\

Now we shall study $Q$  near the boundary, $r\to\infty$. We Taylor expand $Q$ by using the fact that $_2F_1\left[a,b,c,0\right]=1$ and (\ref{deriv}), which result in
\begin{equation}
\left. {_2F_1}\left[\frac{u+\nu+1}{2}, \frac{u+\nu+2}{2}; \nu+3/2;\frac{1}{x^2}\right]\right|_ {\frac{1}{x^2}=0}=1+\frac{1}{x^2}\frac{(u+\nu+1)(u+\nu+2)}{4(\nu+3/2)}
\end{equation}
we find that
\begin{align}
Q^u_{\nu}(x)&\simeq\frac{\sqrt{\pi}\Gamma[\nu+u+1]}{2^{\nu+1}\Gamma[\nu+3/2]}\frac{1}{x^{\nu+1}}
\left(-\frac{u}{2x^2}+1\right)~{_2F_1}\left[\frac{u+\nu+1}{2}, \frac{u+\nu+2}{2}; \nu+3/2;\frac{1}{x^2}\right]\nonumber \\ & \simeq\frac{\sqrt{\pi}\Gamma[\nu+u+1]}{2^{\nu+1}\Gamma[\nu+3/2]}\frac{1}{x^{\nu+1}}
\left(1-\frac{u}{2x^2}+\frac{1}{x^2}\frac{(u+\nu+1)(u+\nu+2)}{4(\nu+3/2)}\right)
\end{align}
We expect to find the asymptotic behavior (\ref{fluctuations}), that is,
 $\phi(r\to\infty)=\frac{m}{r^{\Delta_-}}+\frac{c}{r^{\Delta_+}}+\ldots$. Here, the correct powers of $r$ are
found by

\begin{equation}
x^{\nu+1}r^{\frac{1}{2}}\sim r^{2\nu+\frac{5}{2}}=r^{\frac{3}{2}+\sqrt{\frac{50f^2-23}{4(1+2f^2)}}}\sim r^{\Delta_+}
\end{equation}
\begin{equation}
x^{-\nu}r^{\frac{1}{2}}\sim r^{\frac{3}{2}-\sqrt{\frac{50f^2-23}{4(1+2f^2)}}}\sim r^{\Delta_-}
\end{equation}
where the extra factor of $r$ comes from definition of $\phi$ in (\ref{solutionofeqmo}). \\
The first asymptotic term (with $m$), which decays slower, always comes from $P$.

\begin{equation}
m~=~\frac{\Gamma[1+2\nu]}{\Gamma[1-u+\nu]\Gamma[1+\nu]}\left(\frac{1}{2}\right)^{\nu}\left({\sqrt{1+2f^2}\over 1+f^2} \frac{1}{b}\right)^\nu~c_1
\end{equation}
The ${\sqrt{1+2f^2}\over 1+f^2} \frac{1}{b}$ term is coming from the fact that $\phi$ is defined in terms of $r$ but the Legendre Functions are in terms of $x$.

Depending on the sign of $\nu$, the second asymptotic term (with $c$) may come from either $Q$ or from both of $P$ and $Q$.
If $\nu>0$, the second term in $P$ goes at most like $x^{\nu-2}$ and first term in $Q$ goes like $x^{-\nu-1}$,
because $\nu \leq\frac{1}{4}$, the term coming from $Q$ is the dominant one.
(We recall that
$\Delta_+=2\nu+\frac{5}{2}$ and $\Delta_-=\frac{1}{2}-2\nu$.)

Looking for the coefficients, one finds that
\begin{equation}
c=-2^u \frac{\Gamma\left[{u+v+1\over2}\right]\Gamma\left[{u+v+2\over2}\right]}{\Gamma[u]\Gamma[1-u]\Gamma\left[\nu+{3\over2}\right]}\left({1+f^2\over\sqrt{1+2f^2}}b\right)^{\nu+1}~c_1
\end{equation}

then

\begin{align}
{c\over m}={-2^{u+\nu}\left(\frac{1+f^2}{\sqrt{1+2f^2}}b\right)^{2\nu+1}
{ \frac{{\Gamma[1-u+\nu]~\Gamma[1+\nu]}~\Gamma\left[{u+v+1\over2}\right]
\Gamma\left[{u+v+2\over2}\right]}{{\Gamma\left[1+2\nu\right]}~\Gamma[u]~\Gamma[1-u]~\Gamma\left[\nu+{3\over2}\right]}
}}
\end{align}

If $\nu<0$, we use the expansion we found for $P$ and $Q$ at $x=\infty$, which results in

\begin{equation}
c~=~2^{\nu}\frac{\Gamma[1-2\nu]}{\Gamma[1-\nu]~\Gamma[1-\nu-u]}\frac{(v+u)}{2\nu+1}~c_1~-2^u \frac{\Gamma\left[{u+v+1\over2}\right]
~\Gamma\left[{u+v+2\over2}\right]}{\Gamma[u]~\Gamma[1-u]~\Gamma\left[\nu+{3\over2}\right]}~c_1
\end{equation}

then
\begin{equation}\label{coverm}
{c\over m}={(\frac{1+f^2}{\sqrt{1+2f^2}})^{2\nu+1}\left[2^{\nu}\frac{\Gamma[1-2\nu]}{\Gamma[1-\nu]\Gamma[1-\nu-u]}\frac{(v+u)}{2\nu+1}-2^u \frac{\Gamma\left[{u+v+1\over2}\right]\Gamma\left[{u+v+2\over2}\right]}{\Gamma[u]\Gamma[1-u]\Gamma\left[\nu+{3\over2}\right]}\right]
\over\frac{\Gamma[1+2\nu]}{\Gamma[1-u+\nu]\Gamma[1+\nu]}(\frac{1}{2})^{\nu}}
\end{equation}


\begin{thebibliography}{99}



\bibitem{maldacena} J.~M.~Maldacena, Adv.~Theor.~Math.~Phys.~{\bf 2},
  231 (1998); S.~S.~Gubser, I.~R.~Klebanov, A.~M.~Polyakov,
  Phys.~Lett.~{\bf B428}, 105 (1998); E.~Witten,
  Adv.~Theor.~Math.~Phys.~{\bf 2}, 253 (1998)253291.

\bibitem{hartnoll}S.~Hartnoll, Class.~Quant.~Grav.~{\bf 26}, 224002
  (2009) arXiv:0903.3246; M.~Rangamani, Class.~Quant.~Grav.~{\bf 26},
  224003 (2009) arXiv:0905.4352; J.~McGreevy, arXiv:0909:0518.


\bibitem{Semenoff:1984dq} G.~W.~Semenoff,
Phys.\ Rev.\ Lett.\
  {\bf 53} 2449 (1984).
  \bibitem{graphene} A.~K.~Geim, K.~S.~Novoselov,
Nat. Mater. {\bf 6} 183 (2007);
K.~Novoselov, Nature Materials {\bf 6} 720 - 721 (2007);
M.~I.~Katsnelson, Materials Today {\bf 10} 20 (2007);
  G.~W.~Semenoff,
  Phys.\ Scripta {\bf T146}, 014016 (2012)
  [arXiv:1108.2945 [hep-th]].


.

\bibitem{kane}C.~L.~Kane, E.~J.~Mele, Phys.~Rev.~Lett.~{\bf 95},146802 (2005);
B.~A.~Bernevig, T.~L.~Hughes, S.-C.~Zhang, Science {\bf 314}  1757-1761 (2006);
L.~Fu, C.~L.~Kane, Phys.~Rev.~B {\bf 76} 045302 (2007); C.~Nayak, S.~H.~Simon, A.~Stern,
 M.~Freedman, and S.~Das Sarma, Rev.~Mod.~Phys.~{\bf 80},
 1083 (2008).

\bibitem{balents}L.\ Balents, M.\ P.\ A.\ Fisher, C.\ Nayak, Int.\ J.\ Mod.\ Phys.\  {\bf 10}, 1033 (1998); M.~Franz, Z.~Te$\breve{s}$anovi$\acute{c}$, Phys.~Rev.~Lett.~{\bf 87}, 257003 (2001);
    I.~Herbut, Phys.~Rev.~{\bf B66}, 094504 (2002).


 \bibitem{dunne}J.~Ruostekoski, G.V.~Dunne, J.~Javanainen, Phys.~Rev.~Lett.~{\bf 88}, 180401 (2002);
 S.-L.~Zhu1, B.~Wang, L.-M.~Duan, Phys.~Rev.~Lett.~{\bf 98}, 260402 (2007); L.~Lepori, G. Mussardo and A. Trombettoni,
arXiv:1004.4744 [hep-th]

\bibitem{Davis:2011gi}
  J.~L.~Davis, H.~Omid and G.~W.~Semenoff,
  ``Holographic Fermionic Fixed Points in d=3,''
  JHEP {\bf 1109}, 124 (2011)
  [arXiv:1107.4397 [hep-th]].

\bibitem{klimenko}K.G.~Klimenko,
Theor.~Math.~Phys.~{\bf 89}, 1161 (1992).

\bibitem{cat1}V.P.~Gusynin, V.A.~Miransky, I.A.~Shovkovy,
Phys.~Rev.~Lett.~{\bf 73} (1994) 3499
[hep-ph/9405262].

\bibitem{cat2}V.P.~Gusynin, V.A.~Miransky, I.A.~Shovkovy,
Phys.~Rev.~{\bf D 52} (1995) 4718 [hep-th/9407168].



 \bibitem{Semenoff:1998bk}
  G.~W.~Semenoff, I.~A.~Shovkovy and L.~C.~R.~Wijewardhana,
Mod.\ Phys.\ Lett.\  A {\bf 13}, 1143 (1998)
  [arXiv:hep-ph/9803371].

\bibitem{Semenoff:1999xv}
  G.~W.~Semenoff, I.~A.~Shovkovy and L.~C.~R.~Wijewardhana,
  Phys.\ Rev.\  D {\bf 60}, 105024 (1999)
  [arXiv:hep-th/9905116].

\bibitem{Semenoff:2011ya}
  G.~W.~Semenoff and F.~Zhou,
JHEP {\bf 1107}, 037 (2011)
  [arXiv:1104.4714].


\bibitem{Gorbar:2011kc}
  E.~V.~Gorbar, V.~P.~Gusynin, V.~A.~Miransky and I.~A.~Shovkovy,
  [arXiv:1105.1360].

\bibitem{Gamayun:2012qr}
  O.~V.~Gamayun, E.~V.~Gorbar and V.~P.~Gusynin,
  [arXiv:1206.2266].


\bibitem{Shovkovy:2012zn}
  I.~A.~Shovkovy,


\bibitem{Filev:2009xp}
  V.~G.~Filev, C.~V.~Johnson and J.~P.~Shock,
  JHEP {\bf 0908}, 013 (2009)
  [arXiv:0903.5345 [hep-th]].



\bibitem{Myers:2008me}
  R.~C.~Myers and M.~C.~Wapler,
  JHEP {\bf 0812}, 115 (2008)
  [arXiv:0811.0480 [hep-th]].



\bibitem{Evans:2008nf}
  N.~Evans and E.~Threlfall,
  Phys.\ Rev.\  D {\bf 79} (2009) 066008
  [arXiv:0812.3273 [hep-th]].

\bibitem{Filev:2009ai}
  V.~G.~Filev,
  JHEP {\bf 0911}, 123 (2009)
  [arXiv:0910.0554 [hep-th]].

\bibitem{Evans:2010iy}
  N.~Evans, A.~Gebauer, K.~Y.~Kim and M.~Magou,
  JHEP {\bf 1003}, 132 (2010)
  [arXiv:1002.1885 [hep-th]].

\bibitem{Jensen:2010ga}
  K.~Jensen, A.~Karch, D.~T.~Son and E.~G.~Thompson,
  Phys.\ Rev.\ Lett.\  {\bf 105}, 041601 (2010)
  [arXiv:1002.3159 [hep-th]].



\bibitem{Evans:2010hi}
  N.~Evans, A.~Gebauer, K.~Y.~Kim and M.~Magou,
Phys.~Lett.~{\bf B698}, 91 (2011).
  [arXiv:1003.2694 [hep-th]].

\bibitem{pal}S.~S.~Pal, 
Phys.~Rev.~{\bf D82}, 086013 (2010). [arXiv:1006.2444v4 [hep-th]].


\bibitem{Evans:2010new}
 N.~Evans, K.~Jensen, K.-Y.~Kim, 
 Phys.~Rev.~{\bf D82} 105012 (2010)[arXiv:1008.1889  [hep-th]].


\bibitem{Bolognesi:2011un}
  S.~Bolognesi and D.~Tong,
  arXiv:1110.5902 [hep-th].

\bibitem{Bolognesi:2012pi}
  S.~Bolognesi, J.~N.~Laia, D.~Tong and K.~Wong,
  JHEP {\bf 1207}, 162 (2012)
  [arXiv:1204.6029 [hep-th]].


\bibitem{Grignani:2012jh}
  G.~Grignani, N.~Kim and G.~W.~Semenoff,
  arXiv:1203.6162 [hep-th].

\bibitem{Preis:2010cq}
  F.~Preis, A.~Rebhan and A.~Schmitt,
  JHEP {\bf 1103}, 033 (2011)
  [arXiv:1012.4785 [hep-th]].

\bibitem{Davis:2011am}
  J.~L.~Davis and N.~Kim,
  JHEP {\bf 1206}, 064 (2012)
  [arXiv:1109.4952 [hep-th]].
\bibitem{rey}S.-J.~Rey, Prog.~Theor.~Phys.~{\bf 177}, 128 (2009).




\bibitem{Bergman:2010gm}
  O.~Bergman, N.~Jokela, G.~Lifschytz and M.~Lippert,
  JHEP {\bf 1010}, 063 (2010)
  [arXiv:1003.4965 [hep-th]].

\bibitem{Jokela:2010nu}
  N.~Jokela, G.~Lifschytz and M.~Lippert,
  JHEP {\bf 1102}, 104 (2011)
  [arXiv:1012.1230 [hep-th]].
\bibitem{Jokela:2011eb}
  N.~Jokela, M.~Jarvinen and M.~Lippert,
  JHEP {\bf 1105}, 101 (2011)
  [arXiv:1101.3329 [hep-th]].




\bibitem{Bergman:2011zz}
  O.~Bergman, N.~Jokela, G.~Lifschytz and M.~Lippert,
  Fortsch.\ Phys.\  {\bf 59}, 734 (2011).


\bibitem{Bergman:2011rf}
  O.~Bergman, N.~Jokela, G.~Lifschytz and M.~Lippert,
  JHEP {\bf 1110}, 034 (2011)
  [arXiv:1106.3883 [hep-th]].

\bibitem{Jokela:2011sw}
  N.~Jokela, M.~Jarvinen and M.~Lippert,
  JHEP {\bf 1201}, 072 (2012)
  [arXiv:1107.3836 [hep-th]].

\bibitem{Jokela:2012vn}
  N.~Jokela, G.~Lifschytz and M.~Lippert,
  JHEP {\bf 1205}, 105 (2012)
  [arXiv:1204.3914 [hep-th]].




\bibitem{voz}J.~Gonzalez, F.~Guinea. M.~A.~H.~Vozmediano, Nucl.
  Phys. B. 424, 595 (1994); Phys. Rev. B 59, R2474 (1999);
  E.~V.~Gorbar, V.~P.~Gusynin, V.~A.~Miransky, Phys.\ Rev.\ D {\bf
    64}, 105028 (2001); E.~V.~Gorbar, V.~P.~Gusynin, V.~A.~Miransky,
  I.~A.~Shovkovy, Phys.~Rev.~{\bf B66}, 045108 (2002); D.E.~Sheehy,
  J.~Schmalian, Phys. Rev. Lett. 99, 226803 (2007); O.~Vafek,
  M.~J.~Case, Phys. Rev. B 77, 033410 (2008); Igor F.~Herbut, Vladimir
  Juricic', Oskar Vafek, Phys.~Rev.~Lett.~{\bf 100} (2008) 046403;
  Vladimir Juricic', Igor F.~Herbut, Gordon W.~Semenoff, Phys.Rev.B
  80, 081405, (2009); Igor F.~Herbut, Vladimir Juricic', Oskar Vafek,
  Phys.Rev.B 82, 235402, (2010).


\bibitem{kp}D.~Kutasov, J.~Lin, A.Parnachev,   arXiv:1107.2324 [hep-th].

\bibitem{Grignani:2012jh}
  G.~Grignani, N.~Kim and G.~W.~Semenoff,
  ``D3-D5 Holography with Flux,''
  arXiv:1203.6162 [hep-th].

\bibitem{Klebanov:1999tb}
  I.~R.~Klebanov and E.~Witten,
  Nucl.\ Phys.\  B {\bf 556}, 89 (1999)
  [arXiv:hep-th/9905104].

\bibitem{fradkin}E.~S.~Fradkin, Proc. of the Lebedev Physics Institute, 29, 6 (1965).

\bibitem{special} George~E.~Andrews, Ranjan Roy, Richard Askey,
  ``Special Functions'' Encyclopedia of Mathematics and its
  Applications, The University Press, Cambridge, 1999.

 \end{thebibliography}
\end{document}